\documentclass[fleqn,usenatbib]{mnras}

\usepackage{newtxtext,newtxmath}
\usepackage[T1]{fontenc}
\usepackage{ae,aecompl}

\usepackage{graphicx}	% Including figure files
\usepackage{amsmath}	% Advanced maths commands
\usepackage{amssymb}	% Extra maths symbols
\usepackage{soul}
\usepackage{multirow}

\title[Iax abundance tomography]{Type Iax SNe as a few-parameter family}

\author[B. Barna et al.]
{Barnab\'as Barna$^{1,2}$\thanks{E-mail: bbarna@titan.physx.u-szeged.hu}, Tam\'as Szalai$^{1}$, Wolfgang E. Kerzendorf$^{2}$, Markus Kromer$^{3,4}$,\newauthor Stuart A. Sim$^{5}$, Mark R. Magee$^{5}$, Bruno Leibundgut$^{2}$
\\
$^{1}$Department of Optics and Quantum Electronics, University of Szeged, Dom ter 9 Szeged, Hungary\\
$^{2}$European Southern Observatory, Karl-Schwarzschild-Strasse 2, 85748 Garching bei M{\"u}nchen, Germany\\
$^{3}$Heidelberger Institut f{\"u}r Theoretische Studien, Schloss-Wolfsbrunnenweg 35, D-69118 Heidelberg, Germany\\
$^{4}$Zentrum f{\"u}r Astronomie der Universit{\"a}t Heidelberg, Institut f{\"u}r Theoretische Astrophysik, Philosophenweg 12, D-69120 Heidelberg, Germany\\
$^{5}$Astrophysics Research Centre, School of Mathematics and Physics, Queen's University Belfast, Belfast BT7 1NN, UK\\
}

\date{Accepted XXX. Received YYY; in original form ZZZ}

\pubyear{2017}

\begin{document}
\label{firstpage}
\pagerange{\pageref{firstpage}--\pageref{lastpage}}
\maketitle

\begin{abstract}
We present direct spectroscopic modeling of five Type Iax supernovae (SNe) with the one dimensional Monte Carlo radiative transfer code \begin{small}TARDIS\end{small}. The abundance tomography technique is used to map the chemical structure and physical properties of the SN atmosphere. Through via fitting of multiple spectral epochs with self-consistent ejecta models, we can then constrain the location of some elements within the ejecta. The synthetic spectra of the best-fit models are able to reproduce the flux continuum and the main absorption features in the whole sample.
We find that the mass fractions of IGEs and IMEs show a decreasing trend toward the outer regions of the atmospheres using density profiles similar to those of deflagration models in the literature. Oxygen is the only element, which could be dominant at higher velocities. The stratified abundance structure contradicts the well-mixed chemical profiles predicted by pure deflagration models.
Based on the derived densities and abundances, a template model atmosphere is created for the SN Iax class and compared to the observed spectra. Free parameters are the scaling of the density profile, the velocity shift of the abundance template, and the peak luminosity. The results of this test support the idea that all SNe Iax can be described by a similar internal structure, which argues for a common origin of this class of explosions.

\end{abstract}

\begin{keywords}
supernovae: general -- line: formation -- line: identification -- radiative transfer
\end{keywords}

%%%%%%%%%%%%%%%%%%%%%%%%%%%%%%%%%%%%%%%%%%%%%%%%%%

%%%%%%%%%%%%%%%%% BODY OF PAPER %%%%%%%%%%%%%%%%%%
\section{Introduction}

Most Type Ia supernovae (SNe Ia) form a one-parameter family often called ``Branch-normal'' \citep{Branch06}. The amount of synthesized $^{56}$Ni determines their peak luminosity, which correlates with the shape of their light curves, making these explosions standardizable candles and ideal for distance measurements on cosmological scales. However, certain thermonuclear white dwarf (WD) explosions do not follow this correlation, and some of their other observable properties are usually also peculiar compared to those of the normal SNe Ia. A special group of these objects, named after the prototypical member of them \citep{Li03}, are called `2002cx'-like SNe, or, as it has come into general use after \citet{Foley13}, Type Iax SNe.

\begin{table*}
	\centering
	\caption{Absolute peak magnitudes in V-band (in r-band for SN 2015H), redshifts, distance moduli and reddening values (both in magnitudes) for each SNe in our sample.}
    \label{tab:sample}
    \begin{tabular}{ccccccc}
		\hline
		Object & $M_\rmn{peak}$ & z & $ \mu$ & $E(B-V)_\rmn{MW} $ & $E(B-V)_\rmn{host} $ &  Paper\\
		\hline
		SN 2011ay & -18.39 $\pm$ 0.18 & 0.021 & 34.69 $\pm$ 0.15 & 0.069 & 0.00 & \begin{tabular}{@{}c@{}}\cite{Szalai15} \end{tabular} \\
        \hline
		SN 2012Z & -18.50 $\pm$ 0.09 & 0.007 & 32.59 $\pm$ 0.09 & 0.036 & 0.07 & \begin{tabular}{@{}c@{}}\cite{Stritzinger14} \end{tabular}\\
        \hline
		SN 2005hk & -18.08 $\pm$ 0.29 & 0.012 & 33.46 $\pm$ 0.27 & 0.019 & 0.09 & \begin{tabular}{@{}c@{}}\cite{Phillips07} \end{tabular}\\
        \hline
        SN 2002cx & -17.62 $\pm$ 0.35& 0.024 & 35.09 $\pm$ 0.32 & 0.034 & 0.00 & \begin{tabular}{@{}c@{}}\cite{Li03}\end{tabular}\\
        \hline
        SN 2015H & -17.27 $\pm$ 0.07& 0.012 & 33.91 $\pm$ 0.07 & 0.048 & 0.00 & \begin{tabular}{@{}c@{}}\cite{Magee16}\end{tabular}\\
        \hline
	\end{tabular}
\end{table*}

\begin{table*}
	\centering
	\caption{Log of the spectra of our sample; the phases are given with respect to B-maximum. The times since explosion ($t_\rmn{exp}$) were fitting parameters for our TARDIS models (see in Sec. \ref{fitting_method}) within $\pm$1.5 day to their estimated values in the referred papers.}
    \label{tab:log}
    \begin{tabular}{cccccc}
		\hline
		MJD & $t_\rmn{exp}$ [days] & Phase [days] & Telescope$\setminus$Instrument & Wavelength [\r{A}] & Paper\\
		\hline
        \multicolumn{6}{c}{SN 2011ay}\\
		\hline
        55642.7 & 10.0 & -3.4 & HET$\setminus$LRS & 4100-10000 & \cite{Szalai15}\\
        55643.7 & 11.0 & -2.4 & HET$\setminus$LRS & 4100-10000 & \cite{Szalai15}\\
        55645.7 & 13.0 & -0.4 & HET$\setminus$LRS & 4100-10000 & \cite{Szalai15}\\
        55647.7 & 15.0 & +1.6 & HET$\setminus$LRS & 4100-10000 & \cite{Szalai15}\\
        55648.7 & 16.0 & +2.6 & Lick$\setminus$Kast & 3350-9850 & \cite{Silverman12}\\
        55650.7 & 18.0 & +4.6 & HET$\setminus$LRS & 4100-10000 & \cite{Szalai15}\\
        55652.7 & 20.0 & +6.6 & Lick$\setminus$Kast & 3400-9700 & \cite{Silverman12}\\
        55655.7 & 23.0 & +9.6 & Lick$\setminus$Kast & 3400-9700 & \cite{Silverman12}\\
        55661.7 & 29.0 & +15.6 & Lick$\setminus$Kast & 3350-10550 & \cite{Silverman12}\\
        55664.7 & 32.0 & +18.6 & HET$\setminus$LRS & 4100-10000 & \cite{Szalai15}\\
        \hline
        \multicolumn{6}{c}{SN 2012Z}\\
		\hline
        55958.2 & 5.4 & -9.2 & Lick$\setminus$Kast & 3400-10000 & \cite{Stritzinger14} \\
        55959.2 & 6.4 & -8.2 & Lick$\setminus$Kast & 3400-10000 & \cite{Stritzinger14} \\
        55960.4 & 7.6 & -7.0 & KAO$\setminus$LOSA & 4100-7900 & \cite{Yamanaka15} \\
        55965.4 & 12.6 & -2.0 & KAO$\setminus$LOSA & 4100-7900 & \cite{Yamanaka15} \\
        55968.5 & 15.7 & +1.1 & KAO$\setminus$LOSA & 4100-7900 & \cite{Yamanaka15} \\
        55973.1 & 20.3 & +5.7 & FLWO$\setminus$FAST & 3400-10000 & \cite{Stritzinger14} \\
		\hline
        \multicolumn{6}{c}{SN 2005hk}\\
		\hline
        53675.2 & 5.3 & -9.3 & FLWO$\setminus$FAST & 3500-7400 & \cite{Blondin12}\\
        53676.2 & 6.3 & -8.3 & Lick$\setminus$KAST & 3300-10400 & \cite{Phillips07}\\
        53678.2 & 8.3 & -6.3 & APO$\setminus$DIS & 3600-9600 & \cite{Phillips07}\\
        53679.4 & 9.5 & -5.1 & Keck$\setminus$LRIS & 3200-9200 & \cite{Phillips07}\\
        53680.1 & 10.2 & -4.4 & APO$\setminus$DIS & 3600-9600 & \cite{Phillips07}\\
        53681.2 & 11.3 & -3.3 & FLWO$\setminus$FAST & 3500-7400 & \cite{Blondin12}\\
        53683.2 & 13.3 & -1.3 & FLWO$\setminus$FAST & 3500-7400 & \cite{Blondin12}\\
        53688.2 & 18.3 & +3.7 & MDM$\setminus$CDSS & 3900-7300 & \cite{Phillips07}\\
        \hline
        \multicolumn{6}{c}{SN 2002cx}\\
		\hline
        52411.2 & 7.2 & -4.0 & FLWO$\setminus$FAST & 3500-7500 & \cite{Li03}\\
        52414.2 & 10.2 & -1.0 & FLWO$\setminus$FAST & 3700-7500 &\cite{Li03}\\
        52427.2 & 23.2 & +12.0 & FLWO$\setminus$FAST & 3700-7500 &\cite{Li03}\\
		\hline
        \multicolumn{6}{c}{SN 2015H}\\
		\hline
        57065.1 & 18.9 & +3.2* & EFOSC2 & 3650-9250 & \cite{Magee16}\\
        57068.2 & 22.0 & +6.3* & EFOSC2 & 3350-10000 & \cite{Magee16}\\
        57072.3 & 26.1 & +10.4* & EFOSC2 & 3350-10000 & \cite{Magee16}\\
        \hline
        \multicolumn{6}{l}{* In case of SN 2015H, r-maximum was used instead of B-maximum.}\\
	\end{tabular}
\end{table*}

The characteristic observational properties of SNe Iax are the low peak luminosities, which extend in a wide range from the extremely faint SN 2008ha \citep[$M_\rmn{V,peak}$ $\sim$ -14 mag; ][]{Valenti09} up to the relatively luminous SN 2011ay \citep[$M_\rmn{V,peak}$ $\sim$ -18.4 mag; ][]{Foley13,Szalai15}. Their photospheric velocities are also significantly lower than those of normal Type Ia SNe, falling typically between 5\,000 and 8\,000 km s$^{-1}$ at maximum light \citep{Foley13}, but objects with expansion velocities of 2\,500 km s$^{-1}$ have also been observed \citep{Foley09,Stritzinger14}. Although a general, but not tight, correlation can be noticed between the luminosity and velocity values, some SNe Iax do not fit into this sequence, like SN 2009ku \citep{Narayan11,Foley13} or SN 2014ck \citep{Tomasella16}.

Beyond the wide scale of peak luminosities, the light curves show further diversity. The rise times of SNe Iax are shorter than the typical values for SNe Ia \citep[18.0 $\pm$ 2.0 days in B-band, as found by][]{Ganeshalingam11}; however, the typically faster decline rates do not seem to correlate with the peak luminosities. The near-infrared light curves do not display a second peak, suggesting strong mixing in the ejecta of SNe Iax \citep{Jha06,Phillips07}.

The early spectra of SNe Iax are dominated by lines of iron-group elements (IGEs). Spectral lines of intermediate mass elements (IME), e.g. Si \begin{small}II\end{small} and Ca \begin{small}II\end{small} are always present, but their strength is far from that observed in normal SNe Ia. High-velocity features have not been observed in spectra of any SNe Iax. At later epochs, both permitted absorption features and forbidden emission lines can be found at optical wavelengths, which is unlike the spectra of other thermonuclear explosions \citep{Jha06}.

Studying SNe Iax has the potential to answer some of the open questions related to the progenitor systems of thermonuclear SNe.
While no progenitor system of normal Type Ia SNe has ever been discovered, \cite{McCully14b} reported a possible source coincident with Type Iax SN 2012Z on a pre-explosion image of HST. The detected luminous blue star could be a potential He star donor to the exploding white dwarf. The existence of such progenitor systems, suggested first by \citet{Foley13}, is supported by detailed binary evolution calculations \citep{Wang13}. However, \cite{Liu15} noted that this kind of system is unlikely to be the progenitor of the majority of SNe Iax, because the long delay time of the single degenerate Chandrasekhar mass ($M_\rmn{Ch}$) models seemingly does not explain the observed number of Type Iax SNe.

The above mentioned observables (i.e. low peak luminosities and photospheric velocities; strong mixing) and the spectral footprints of chemical elements together indicate that deflagrations play a major role in the explosion of Type Iax SNe. Subsonic explosions of carbon-oxygen (CO) $M_\rmn{Ch}$ WDs leaving bound remnants (also referred to as ``failed'' SNe) have been investigated by 3D hydrodynamical simulations \citep{Jordan12,Kromer13}. According to these and other studies \citep{Long14,Magee16}, the synthetic observables of weak pure deflagrations are able to broadly reproduce both the photometric and spectroscopic properties of SNe Iax.

\cite{Fink14} presented a set of 3D hydrodynamical calculations using a multi-spot ignition approach to scale the strength of the deflagration. The peak luminosities of their models N3def, N5def, N10def, and N20def are comparable with the more luminous SNe Iax \citep{Foley13}. The growing number of ignition spots produces more energetic and luminous explosions in accordance with the wide range of the observed luminosities and expansion velocities. In addition, the pure deflagration models of \cite{Fink14} show highly mixed abundance structures with nearly constant mass fractions for each elements.

The main goal of this study is to carry out a comprehensive abundance tomography \citep{Stehle05} analysis involving several different SNe Iax. The sample of objects is chosen to broadly represent the properties of the class. Similar to our pilot study for SN 2011ay \citep{Barna17}, we compare our findings with the results of previous spectral models, as well as with the predictions of the pure deflagration of a Chandrasekhar-mass WD with a bound remnant (the most promising explosion scenario for SNe Iax).

This paper is structured in the following way. 
In Section \ref{sample}, we give an overview of the target sample of our study. In Section \ref{method}, we show how we use the \begin{small}TARDIS\end{small}\footnote{The TARDIS software package available from: \url{https://zenodo.org/record/1292315} .} code \citep{Kerzendorf14} for abundance tomography to simulate the spectral evolution of the objects. In Section \ref{results}, we present the results of our spectral modeling, compare the resulting physical picture of the ejecta with other models and introduce a template for the subclass of SNe Iax. Finally, we summarize our main findings in Section \ref{conclusions}.

\section{SN Iax sample of study} \label{sample}

The adopted technique requires a time sequence. The photosphere assumption of \begin{small}TARDIS\end{small} \citep{Kerzendorf14} is robust only within $\sim$30 days after the explosion (see Sec. \ref{method}), hence only SNe with at least three spectra within this time range were taken into account. As a further constraint, we need reliable information on the host galaxy reddening to fit the emergent luminosity. 

Nearly 60 SNe Iax have been discovered to date, but only a few meet the above-mentioned criteria of our analysis. According to the conditions listed above, five SNe Iax (see Table \ref{tab:sample}) were chosen from open access databases \citep{Yaron12,Guillochon17}. A log of spectroscopic observations used in this study is presented in Table \ref{tab:log}.

The Milky Way galaxy reddening value was adopted from the dust emission map of \cite{Schlafly11} assuming $R_V$ = 3.1 as the extinction coefficient. All spectra are corrected for the sum of galactic and extra-galactic interstellar reddening based on the extinction model of \cite{Fitzpatrick07}.

\subsection{SN 2011ay}

SN 2011ay is one of the most luminous SNe Iax \citep{Foley13} with a peak absolute brightness of $M_\rmn{V,peak}$ $\sim$ -18.4 mag \citep{Szalai15}. It was discovered by the Katzman
Automatic Imaging Telescope (KAIT)/Lick Observatory Supernova
Search (LOSS) programme in NGC 2318 \citep{Blanchard11} at a redshift of $z$ = 0.021. The host galaxy reddening was found to be negligible, since no significant Na D lines were detected.
%The spectroscopic data of SN 2011ay were obtained with the 9.2-m Hobby-Eberly Telescope (HET) Marcario Low Resolution Spectrograph and the 3-m Shane Telescope Kast spectrograph at the Lick Observatory. 

The original spectroscopic data  of SN 2011ay were published by \cite{Silverman12}, while \cite{Szalai15} provided detailed photometric and spectroscopic analysis. Moreover, \cite{Barna17} published an abundance tomography analysis using \begin{small}TARDIS\end{small}; a revision of these results is also part of the current paper.

\subsection{SN 2012Z}
SN 2012Z was also discovered \citep{Cenko12} by the KAIT/LOSS programme. The peak luminosity of SN 2012Z is comparable to that of SN 2011ay; depending on the adopted distance and host galaxy reddening, it may be the most luminous Type Iax SN \citep{Stritzinger14}. The redshift of its host galaxy, NGC 1309 is $z$ = 0.007. The host galaxy reddening was estimated to be $E(B-V)$ = 0.07 mag based on the study of high resolution Na\,\begin{small}I\end{small} $\lambda$5\,890 $\lambda$5\,896 and K\,\begin{small}I\end{small} $\lambda$7\,665 $\lambda$7\,699 line profiles \citep{Stritzinger14}, line strengths of the diffuse interstellar band at 5780 \r{A} \citep{Phillips13}. 

The spectra of SN 2012Z were published by \cite{Yamanaka15} and \cite{Stritzinger14}, who also presented detailed optical spectroscopic analyses.

%The spectra of SN 2012Z were observed with the LOSA/F2 spectrograph mounted on the 1.3-m Araki telescope at the Koyama Astronomical Observatory and with the Lick/Kast spectrograph, published by \cite{Yamanaka15} and \cite{Stritzinger14}, respectively, who also presented detailed optical spectroscopic analyses.

\subsection{SN 2005hk}
SN 2005hk is one of the best observed SNe Iax. It was discovered \citep{Burket05} independently by LOSS and SDSS-II in UGC 272 at a redshift of $z$ = 0.012. The absolute magnitude of SN 2005hk is -18.08 mag in the V-band \citep{Phillips07}. The host galaxy reddening was studied via interstellar polarization and Na D lines \citep{Chornock06} and found to be $E(B-V)$ = 0.09 mag. Different kinds of spectral modeling were carried out by \citet{Phillips07}, \citet{Sahu08} and \citet{Magee17}; while the authors of the former paper used SYNOW \citep{Fisher00}, the latter studies employed the Monte Carlo spectrum synthesis codes of \citet{Mazzali00}, and \citet{Kerzendorf14}, respectively.

\subsection{SN 2002cx}
SN 2002cx is the prototype of the Type Iax SN-class. It was discovered in the galaxy CGCG 044-035 \citep{Wood-Vasey02} at a redshift of $z$ = 0.024. The peak luminosity of SN 2002cx is $M_\rmn{V,peak}$ $\sim$ -17.6 mag \citep{Li03}.
Near-maximum light spectra were first published by \cite{Li03}, who claimed that the object suffered from no or negligible host-galaxy reddening. Detailed spectral analysis regarding the photospheric phase was presented by \cite{Branch04}, while \cite{Jha06} presented late-time spectra.

\subsection{SN 2015H}
SN 2015H was originally discovered \citep{Parker15} by the BOSS program in NGC~3464 ($z$ = 0.012), approximately twenty days after explosion. Thus, no photometric or spectroscopic data were obtained during the pre-maximum phase. An exception is the r-band, in which SN 2015H is well observed and shows a peak absolute magnitude of -17.3 mag. The observables were presented by \cite{Magee16}, who modeled one spectrum with \begin{small}TARDIS\end{small}, and compared their findings with the predictions of deflagration models.

\section{Method}
\label{method}

We use the one dimensional radiative spectral synthesis code \begin{small}TARDIS\end{small} \citep{Kerzendorf14} to perform the abundance tomography technique, which was originally described by \cite{Stehle05}. As the homologously expanding SN ejecta becomes optically thinner, the velocity of the photosphere decreases and observations can probe deeper into the ejecta. Thus, the analysis of a spectral time series allows us to map the different regions of the SN ejecta.

\begin{small}TARDIS\end{small} assumes a sharp photosphere emitting a blackbody continuum, above which the model atmosphere is divided into multiple spherically symmetric shells. Indivisible photon packets representing bundles of photons with the same frequency are sent from the bottom of the computation volume and the code follows their interaction with matter. he chemical abundances and the densities are specified as input parameters in each shell, while the additional physical parameters (radiation temperature, ionization and excitation ratios) are computed by the code iteratively.

The temperature of the photosphere is estimated according to the Stefan-Boltzman law from photospheric velocity and and the emergent luminosity. Note that after each \begin{small}TARDIS\end{small} iteration the statistics of the Monte Carlo packets are used to recalculate the blackbody temperature at the inner boundary to match the value of the emergent luminosity. The radiative temperature in each radial shell is updated according to the Monte Carlo estimators \citep{Kerzendorf14} based on the flight histories of the photon packets. Thus, the temperature-dependent conditions are improved iteratively to a more consistent model.

\begin{figure*}
\centering
\includegraphics[width=15cm]{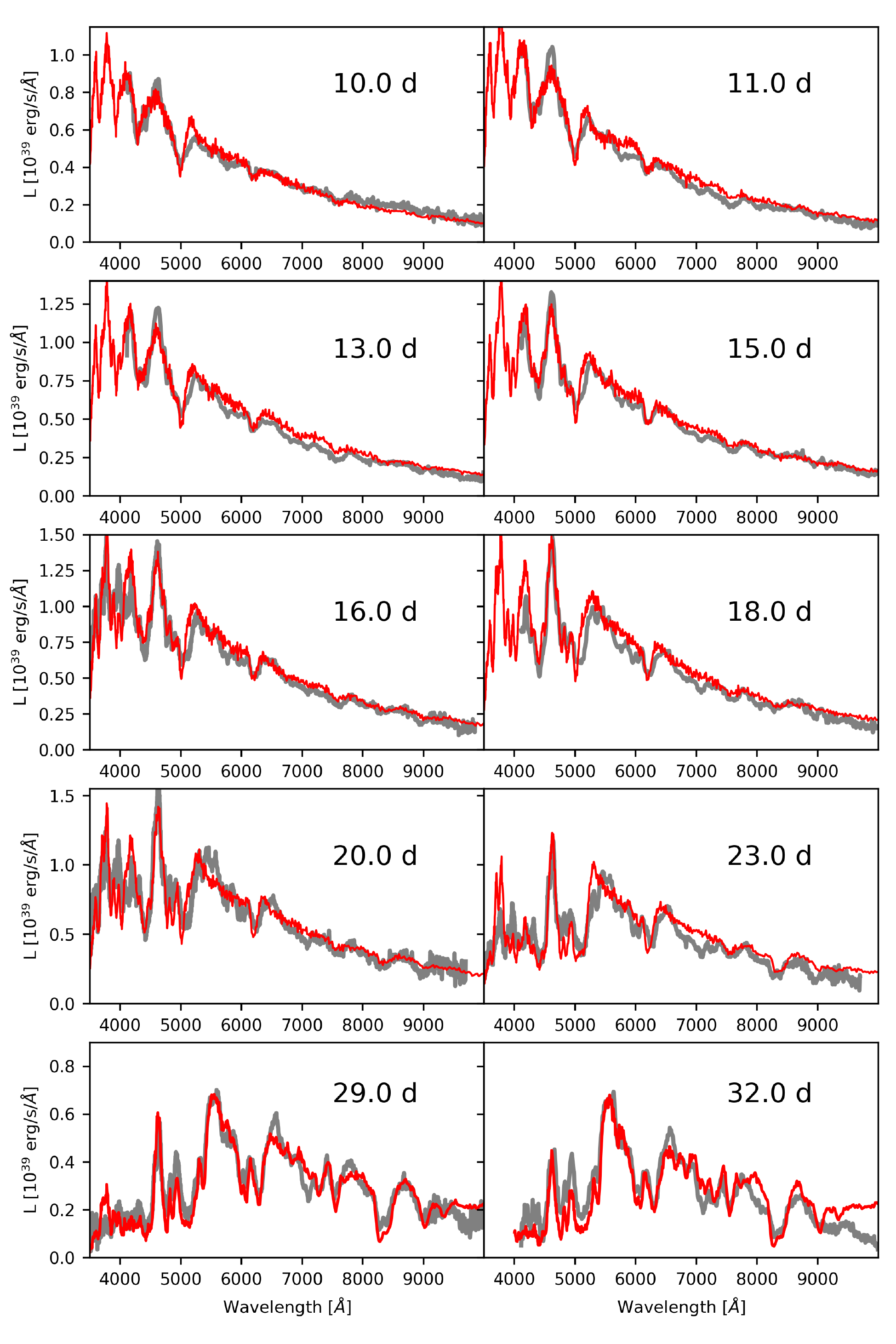}
\caption{\label{fig:sn11ay_spectra} The observed spectra (grey) of SN 2011ay obtained between -3 and +19 days with respect to B-max, compared to our best-fit TARDIS models (red). The numbers show the time since explosion in our best-fit TARDIS models.}
\end{figure*}

\begin{figure*}
\centering
\includegraphics[width=15cm]{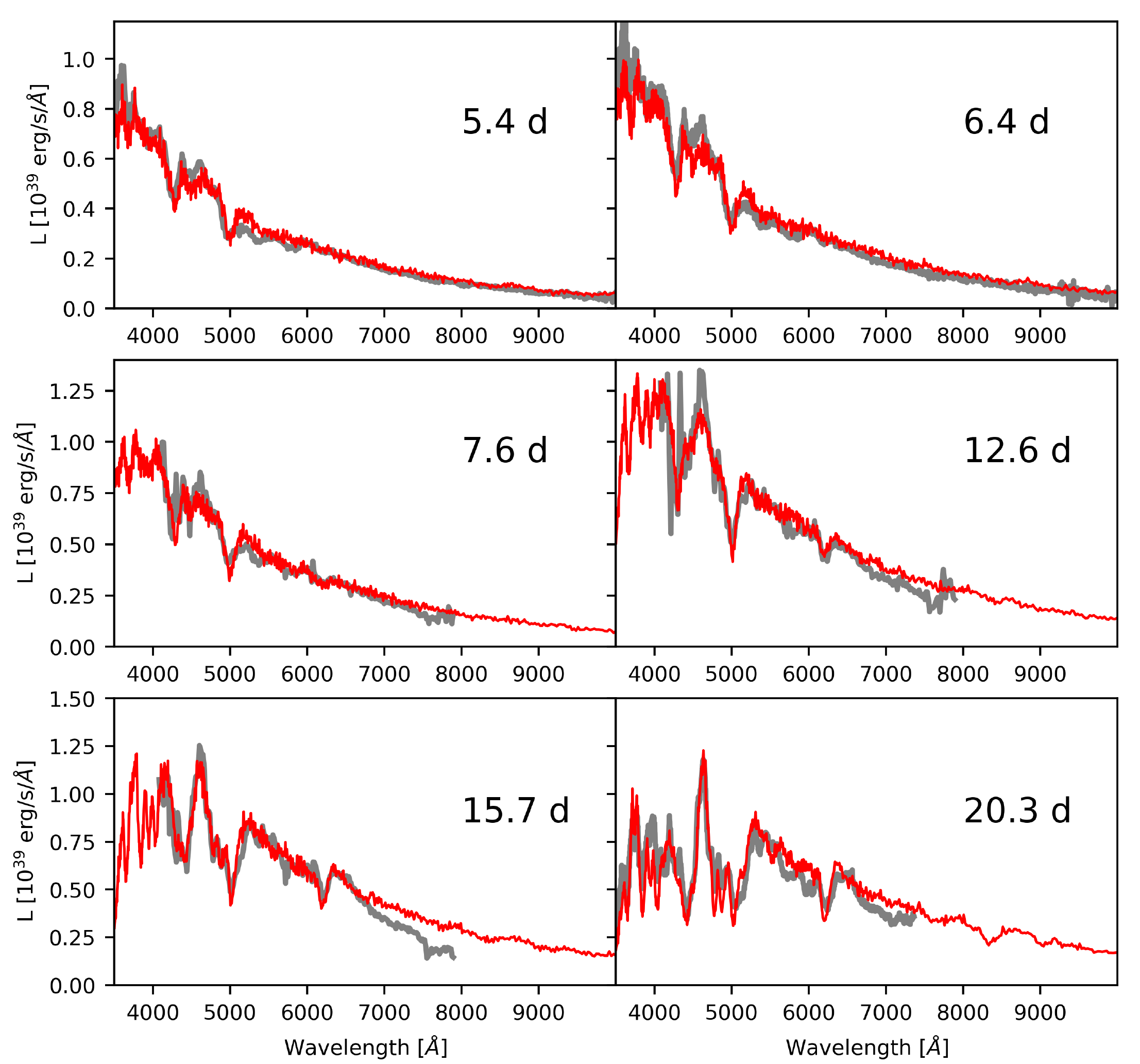}
\caption{\label{fig:sn12Z_spectra} Same as Fig. \ref{fig:sn11ay_spectra} for SN 2012Z obtained between -9 and +6 days with respect to B-max.}
\end{figure*}

\begin{figure*}
\centering
\includegraphics[width=15cm]{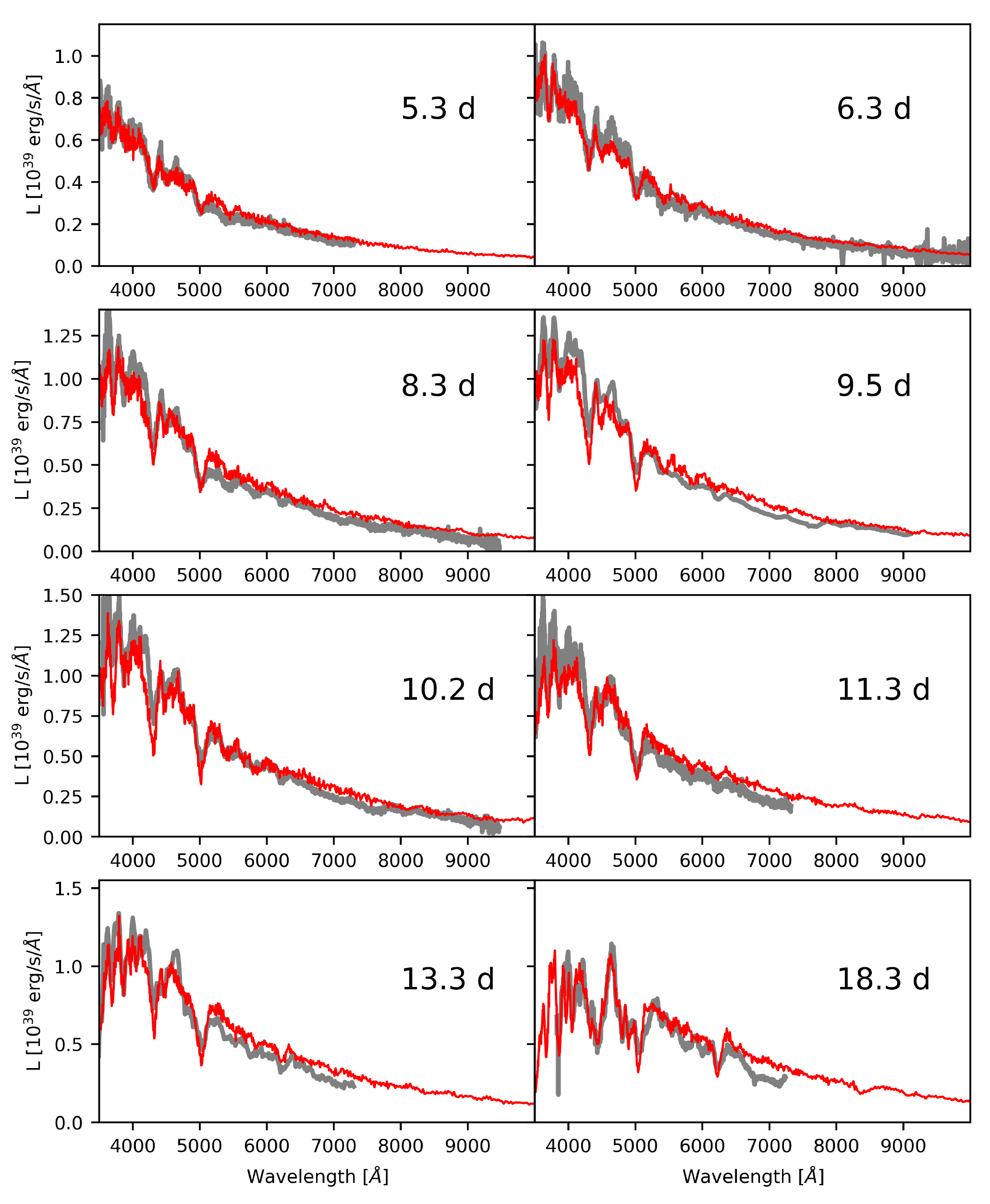}
\caption{\label{fig:sn05hk_spectra} Same as Fig. \ref{fig:sn11ay_spectra} for SN 2005hk obtained between -9 and +4 days with respect to B-max.}
\end{figure*}

\begin{figure}
\centering
\includegraphics[width=\columnwidth]{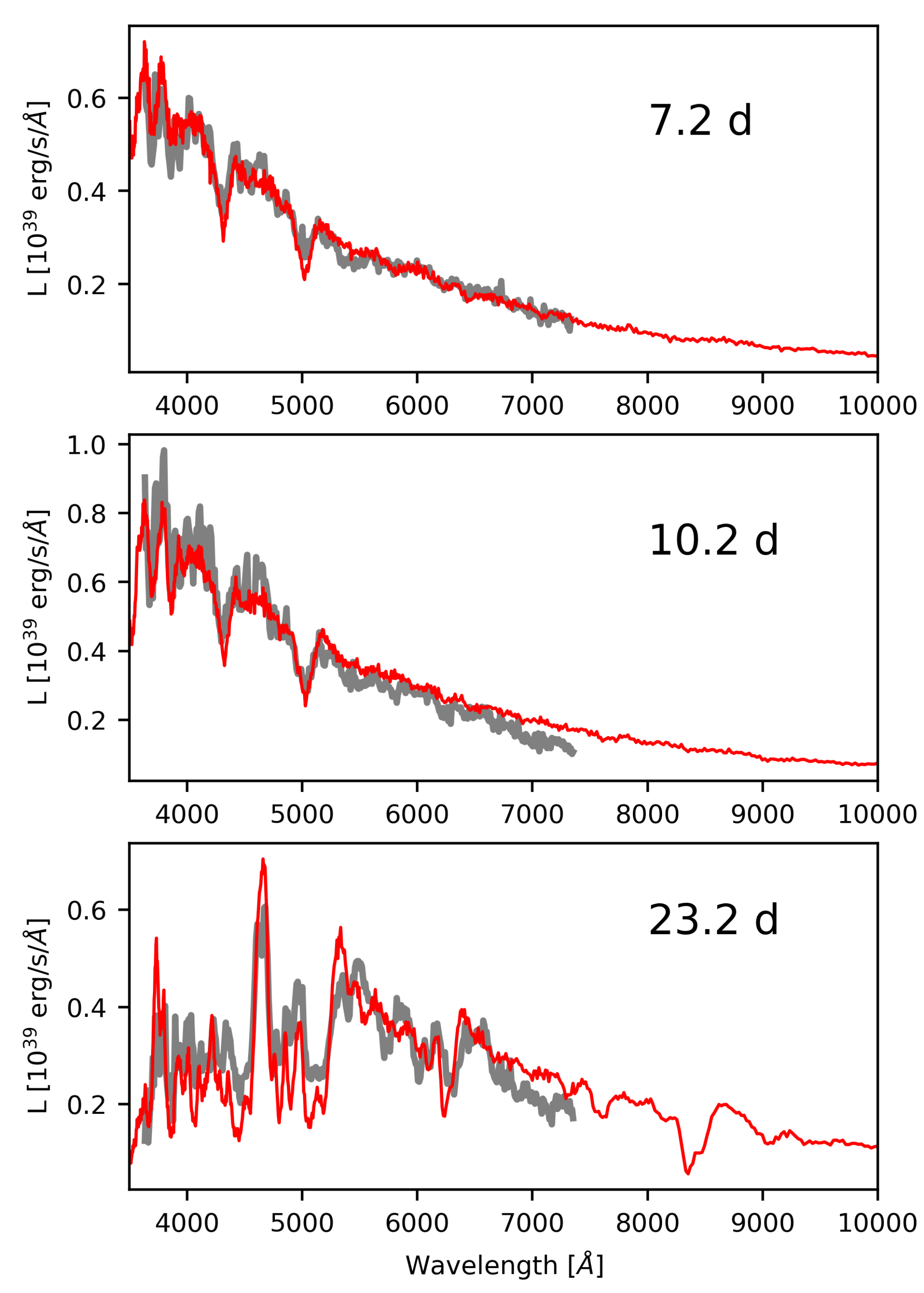}
\caption{\label{fig:sn02cx_spectra} Same as Fig. \ref{fig:sn11ay_spectra} for SN 2002cx obtained between -4 and +12 days with respect to B-max.}
\end{figure}

\begin{figure}
\centering
\includegraphics[width=\columnwidth]{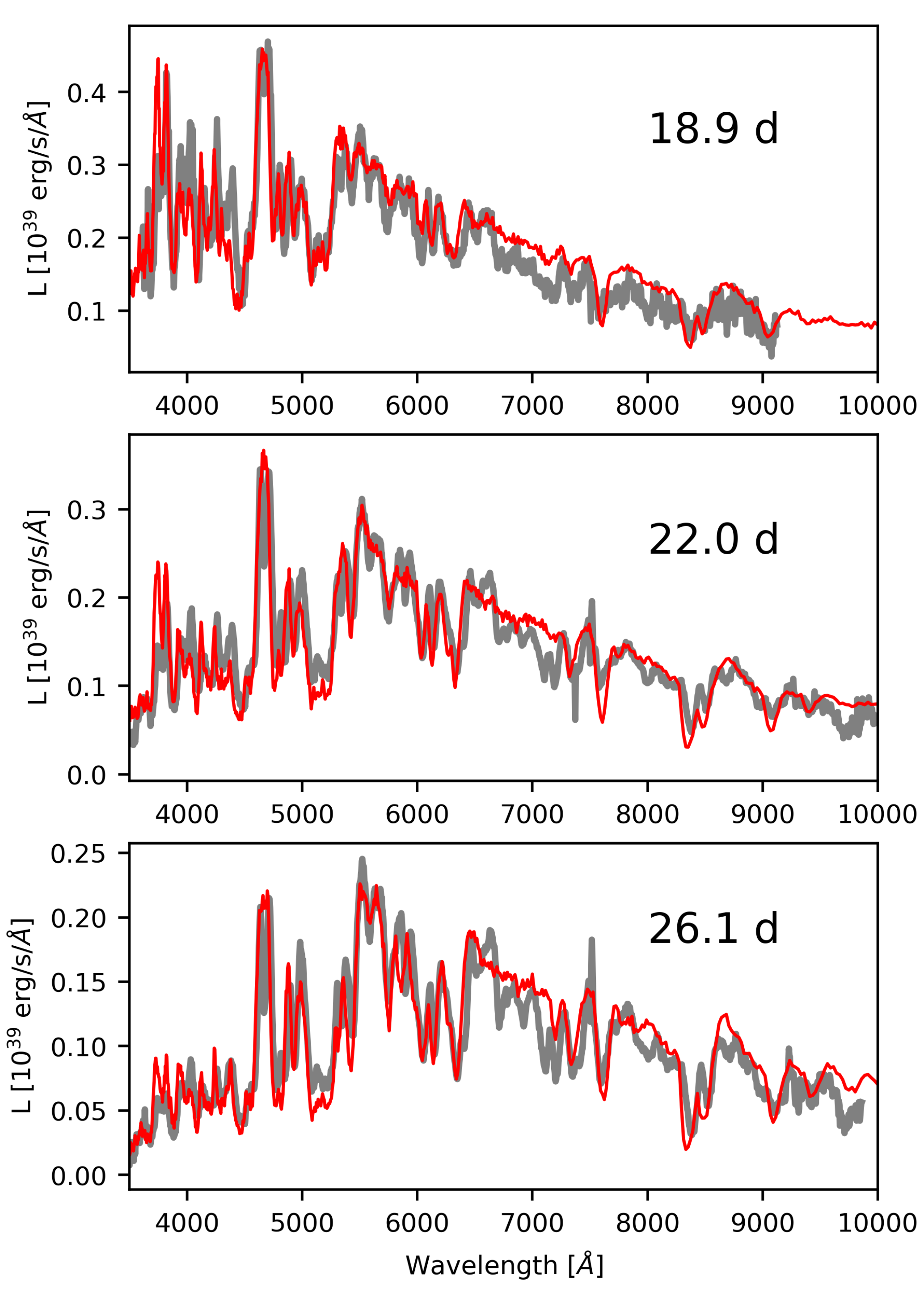}
\caption{\label{fig:sn15H_spectra} Same as Fig. \ref{fig:sn11ay_spectra} for SN 2015H obtained between +3 and +10 days with respect to r-max.}
\end{figure}

\begin{table*}
	\centering
	\caption{The time since explosion, the emergent luminosity and the photospheric velocity parameters of the best-fit TARDIS models. In the rows of the SNe, the central density values (at reference time of 100 s after the explosion) and the locations of the cut-offs in the adopted density functions are shown. The figures show the fitted density profiles of the SNe (solid lines) and the corresponding density functions of the theoretical models (dashed lines)}
    \label{tab:log_tardis}
    \begin{tabular}{ccccccc}
		\hline
		$t_\rmn{exp}$ [days] & $L_\rmn{e}$ [log L$_{\odot}$] & $v_\rmn{phot}$ [km s$^{-1}$] &  &  & $\rho_{\rmn{0}}$ [g cm$^{-3}$] & $v_\rmn{cut}$ [km s$^{-1}$] \\
		\hline
        \multicolumn{3}{c}{SN 2011ay} &  &  & 4.65 & 9\,500 \\
		\hline
         \rule{0pt}{2.5mm}10.0 & 8.94 & 10\,000 & \multicolumn{4}{c}{\multirow{6}{*}{\begin{minipage}{.3\textwidth}\includegraphics[width=\linewidth, height=35mm]{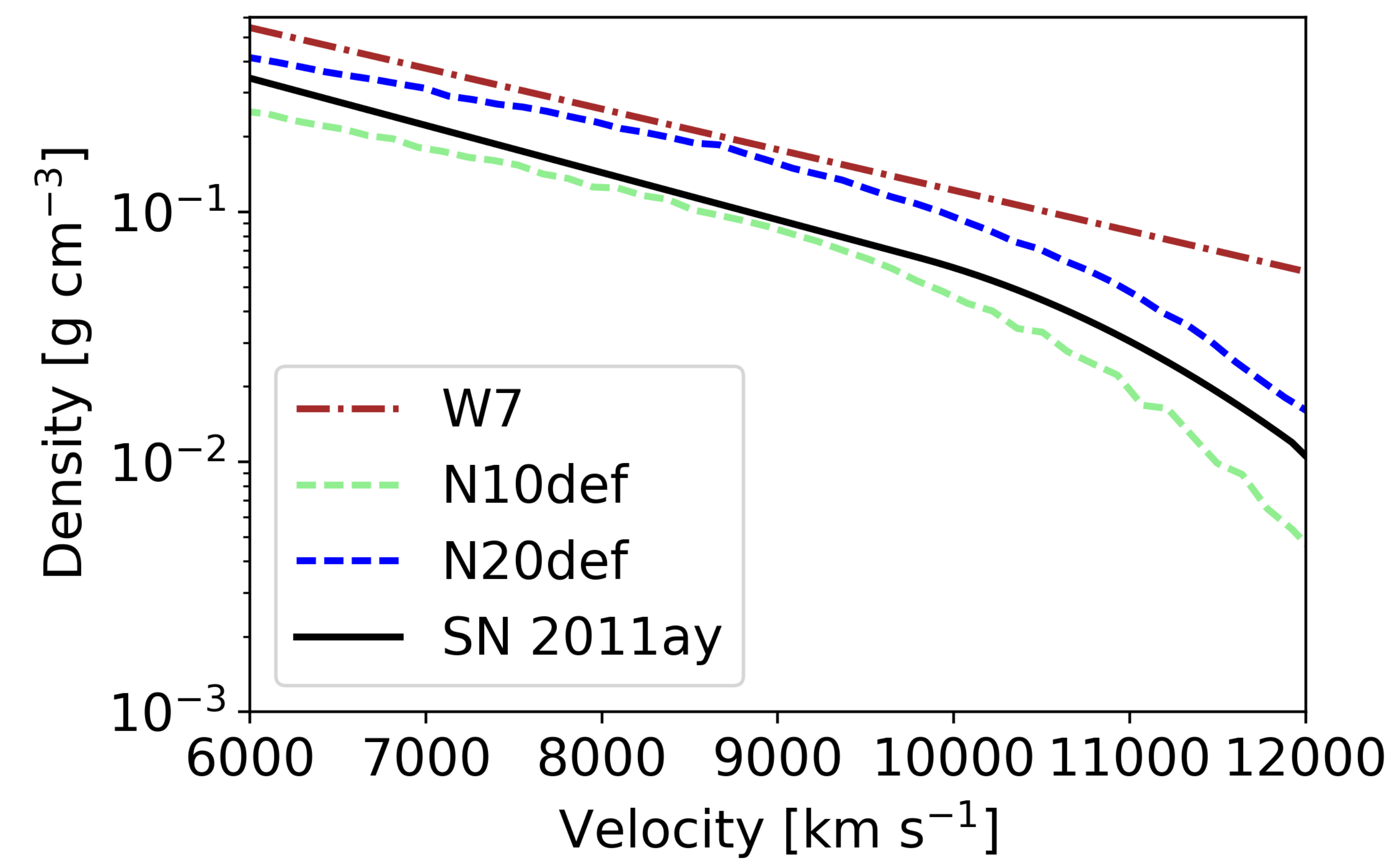}\end{minipage}}}\\
         \rule{0pt}{2.5mm}11.0 & 8.99 & 9\,800\\
         \rule{0pt}{2.5mm}13.0 & 9.04 & 9\,400\\
         \rule{0pt}{2.5mm}15.0 & 9.07 & 9\,200\\
         \rule{0pt}{2.5mm}16.0 & 9.10 & 9\,000\\
         \rule{0pt}{2.5mm}18.0 & 9.11 & 8\,900\\
         \rule{0pt}{2.5mm}20.0 & 9.11 & 8\,500\\
         \rule{0pt}{2.5mm}23.0 & 9.03 & 8\,300\\
         \rule{0pt}{2.5mm}29.0 & 8.86 & 7\,400\\
         \rule{0pt}{2.5mm}32.0 & 8.82 & 6\,800\\
        \hline
        \multicolumn{3}{c}{SN 2012Z} &  &  & 3.80 & 8\,400 \\
		\hline
         \rule{0pt}{2.5mm}5.4 & 8.86 & 10\,700  & \multicolumn{4}{c}{\multirow{6}{*}{\begin{minipage}{.3\textwidth}\includegraphics[width=\linewidth, height=35mm]{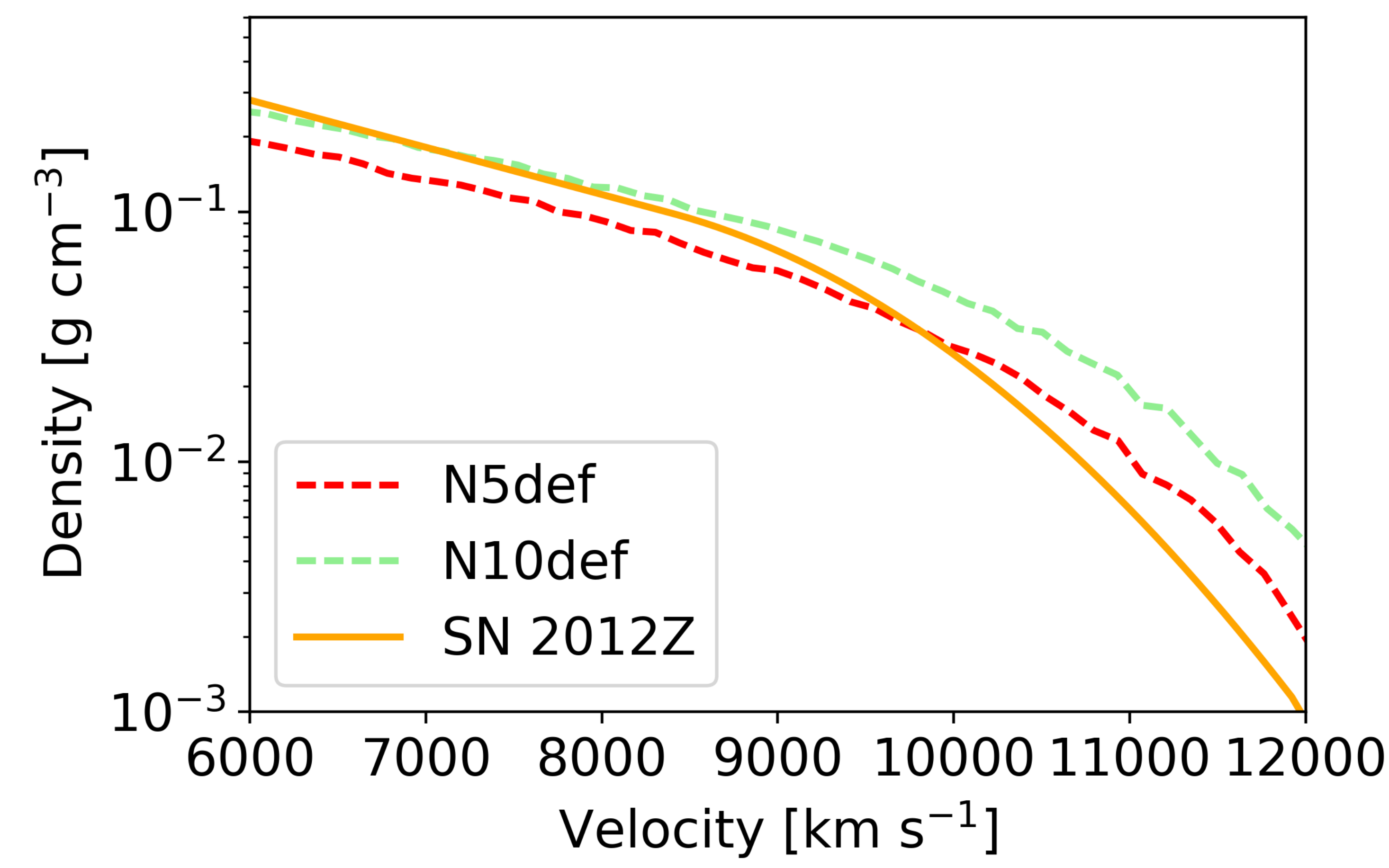}\end{minipage}}}\\
         \rule{0pt}{5mm}6.4 & 8.92 & 10\,300 \\
         \rule{0pt}{5mm}7.6 & 8.94 & 10\,000 \\
         \rule{0pt}{5mm}12.6 & 9.06 & 9\,000 \\
         \rule{0pt}{5mm}15.7 & 9.04 & 8\,500 \\
         \rule{0pt}{5mm}20.3 & 9.00 & 7\,700 \\
		\hline
        \multicolumn{3}{c}{SN 2005hk} &  &  & 2.80 & 7\,100 \\
		\hline
         \rule{0pt}{2.5mm}5.3 & 8.83 & 9300 & \multicolumn{4}{c}{\multirow{8}{*}{\begin{minipage}{.3\textwidth}\includegraphics[width=\linewidth, height=35mm]{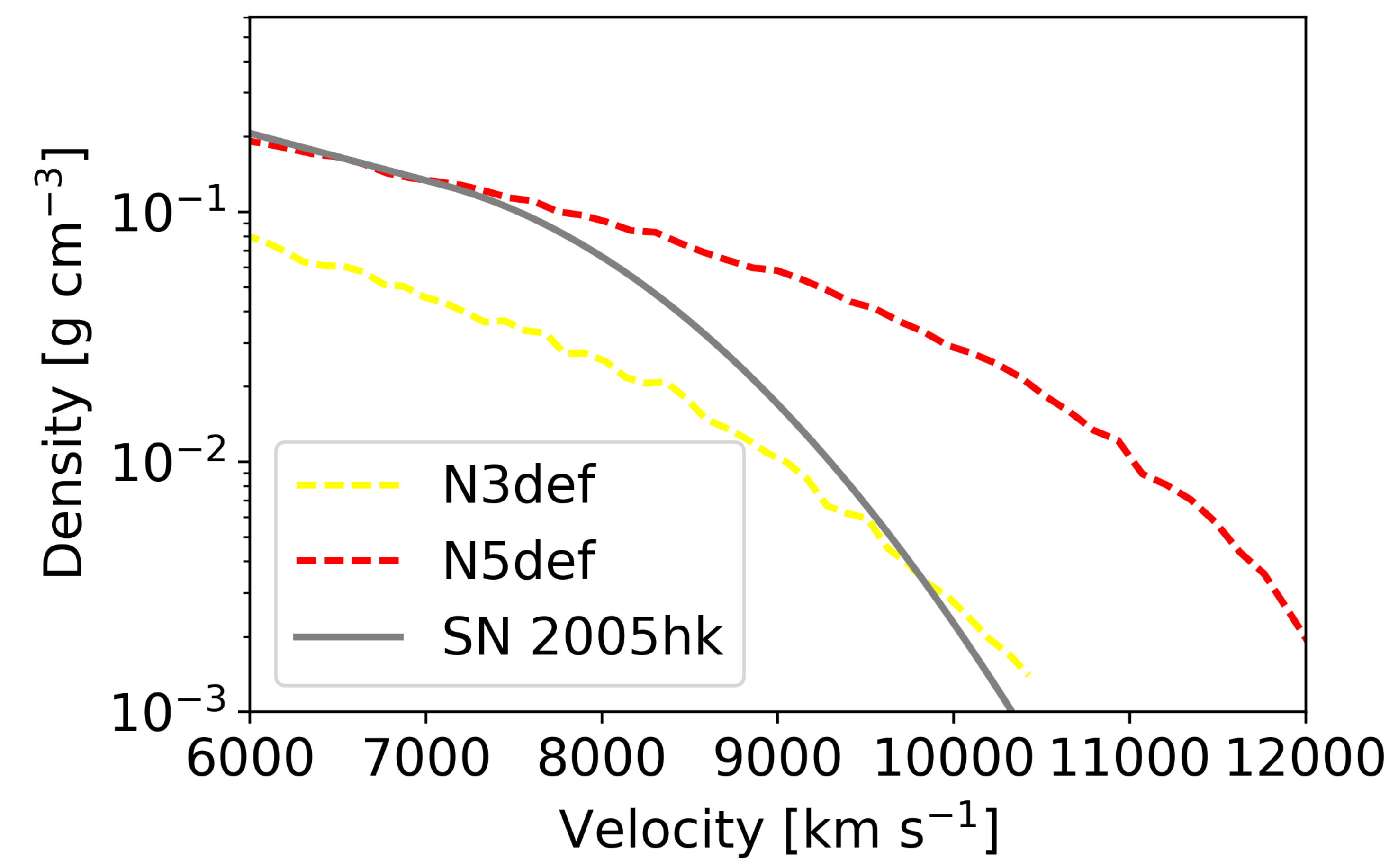}\end{minipage}}}\\
         \rule{0pt}{3.5mm}6.3 & 8.92 & 9\,000\\
         \rule{0pt}{3.5mm}8.3 & 8.99 & 8\,600\\
         \rule{0pt}{3.5mm}9.5 & 9.01 & 8\,200\\
         \rule{0pt}{3.5mm}10.2 & 9.03 & 8\,000\\
         \rule{0pt}{3.5mm}11.3 & 8.99 & 7\,900\\
         \rule{0pt}{3.5mm}13.3 & 9.02 & 7\,600\\
         \rule{0pt}{3.5mm}18.3 & 8.98 & 6\,900\\
        \hline
        \multicolumn{3}{c}{SN 2002cx} &  &  & 1.75 & 7\,000 \\
		\hline
         \rule{0pt}{2.5mm}7.2 & 8.74 & 8200 & \multicolumn{4}{c}{\multirow{3}{*}{\begin{minipage}{.3\textwidth}\includegraphics[width=\linewidth, height=35mm]{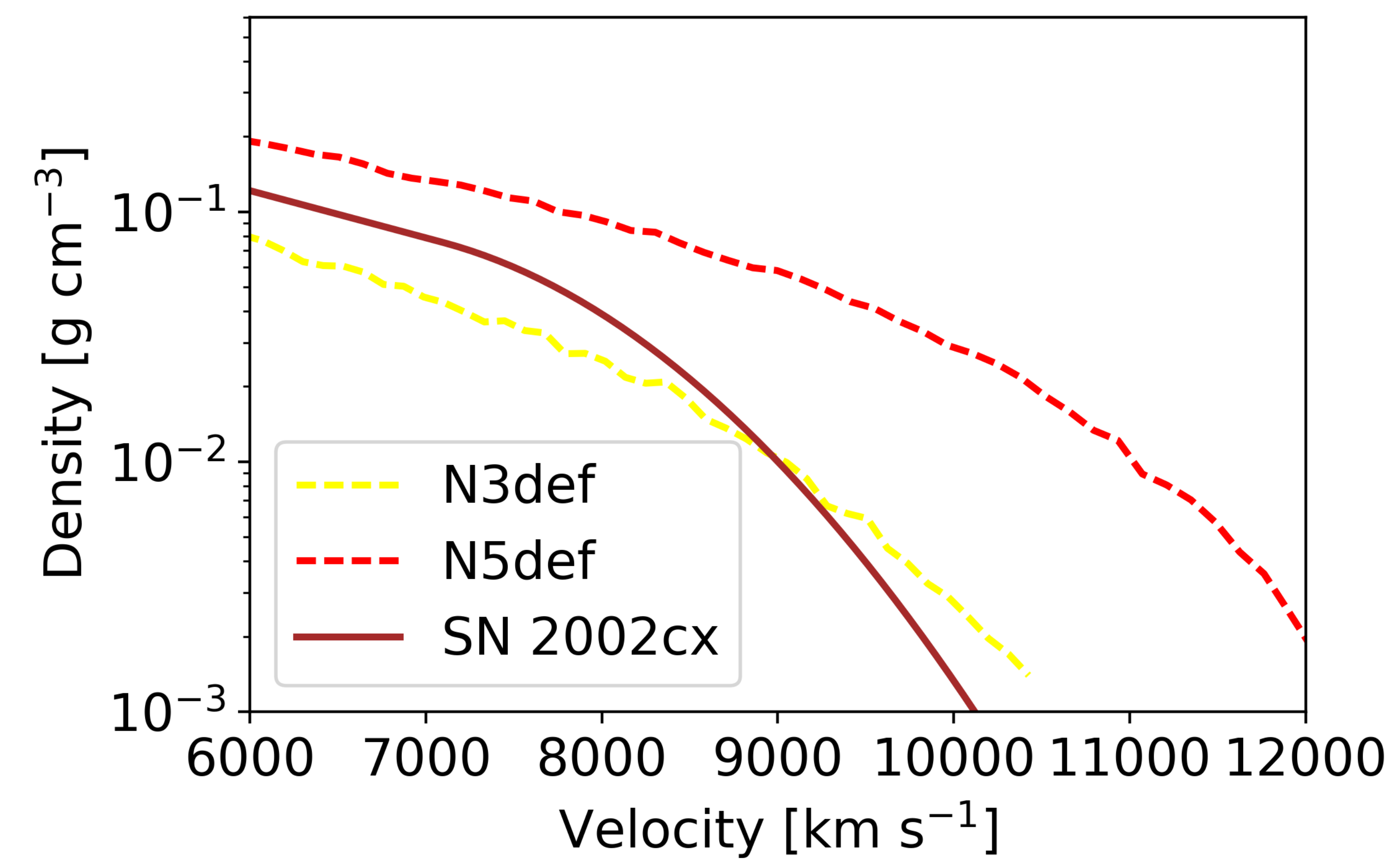}\end{minipage}}}\\
         \rule{0pt}{15mm}10.2 & 8.82 & 7\,500\\
         \rule{0pt}{15mm}23.2 & 8.74 & 5\,600\\
		\hline
        \multicolumn{3}{c}{SN 2015H} &  &  & 1.06 & 7\,100 \\
		\hline
         \rule{0pt}{2.5mm}18.9 & 8.62 & 5800 & \multicolumn{4}{c}{\multirow{3}{*}{\begin{minipage}{.3\textwidth}\includegraphics[width=\linewidth, height=35mm]{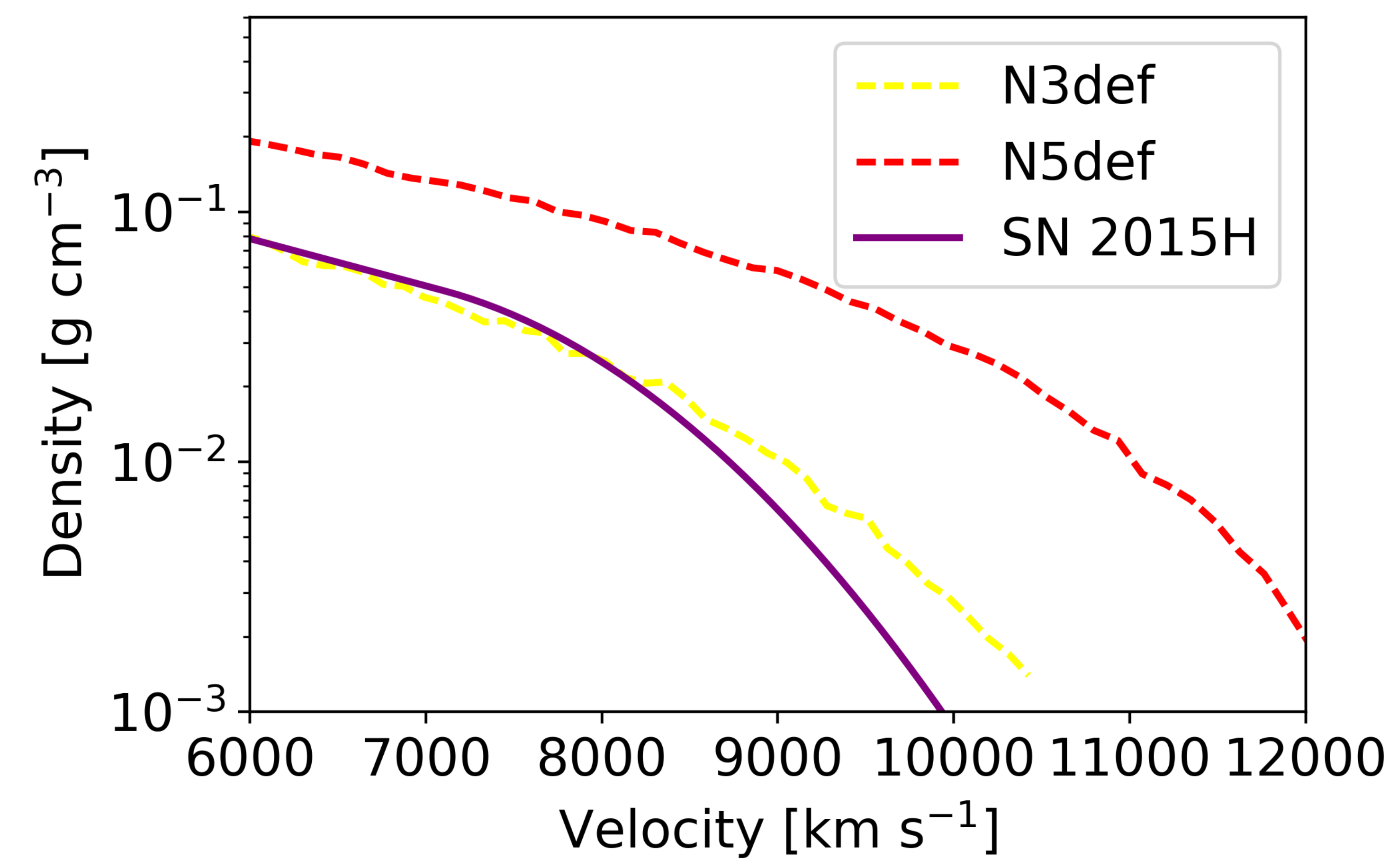}\end{minipage}}}\\
         \rule{0pt}{15mm}22.0 & 8.52 & 5\,500\\
        \rule{0pt}{15mm}26.1 & 8.42 & 5\,100\\
        \hline
	\end{tabular}
\end{table*}

\begin{figure}
\centering
\includegraphics[width=\columnwidth]{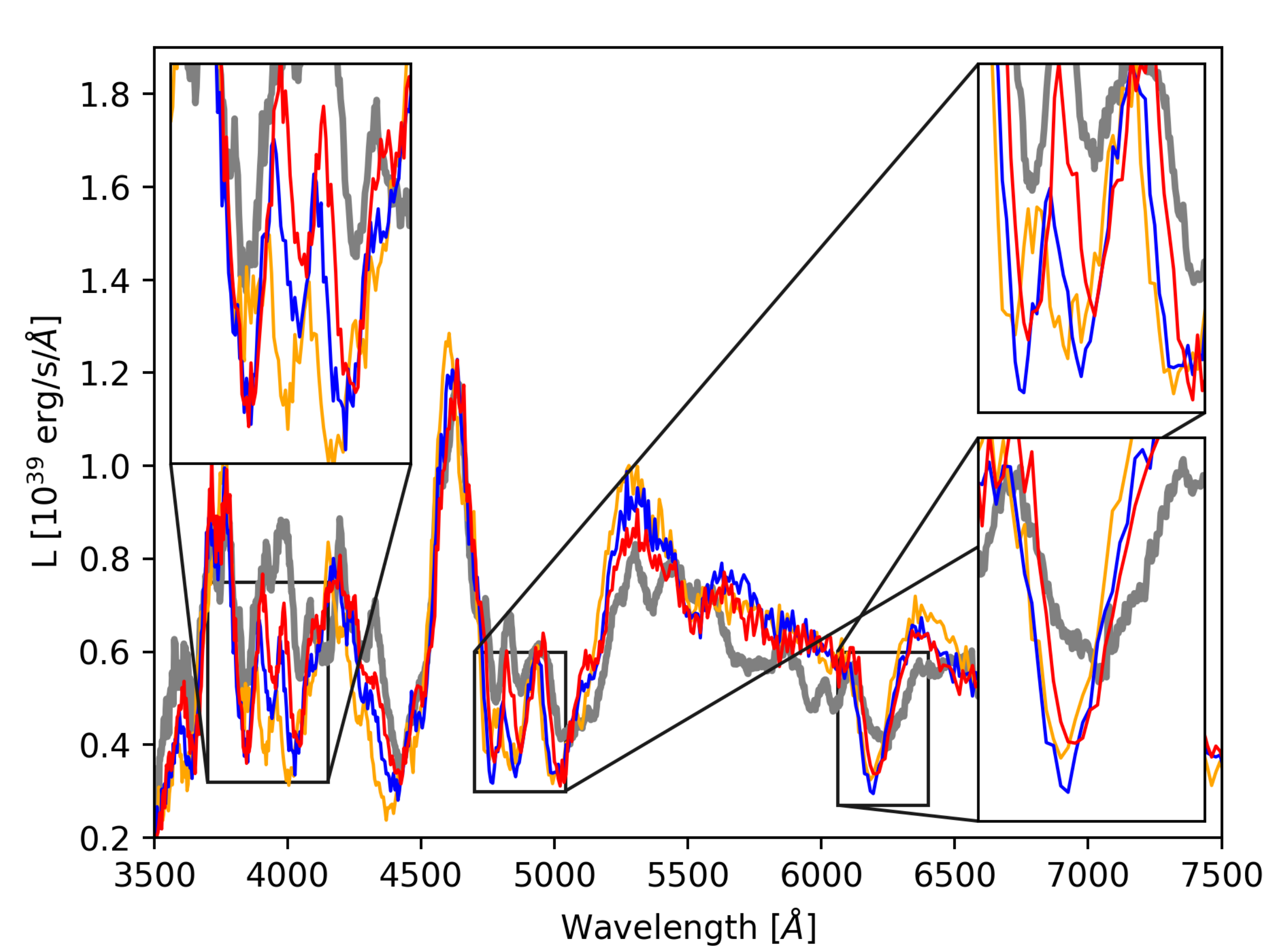}
\caption{\label{fig:density_sn12Z} The spectrum of SN 2012Z at the epoch +1.1 days with respect to B-maximum (grey), the synthetic spectrum from the best-fit  model (red) and the same model with the density profile of the N5def model (blue) and with the density profile without any cut-off at higher velocities (orange).}
\end{figure}

\begin{figure*}
\centering
\includegraphics[width=14cm]{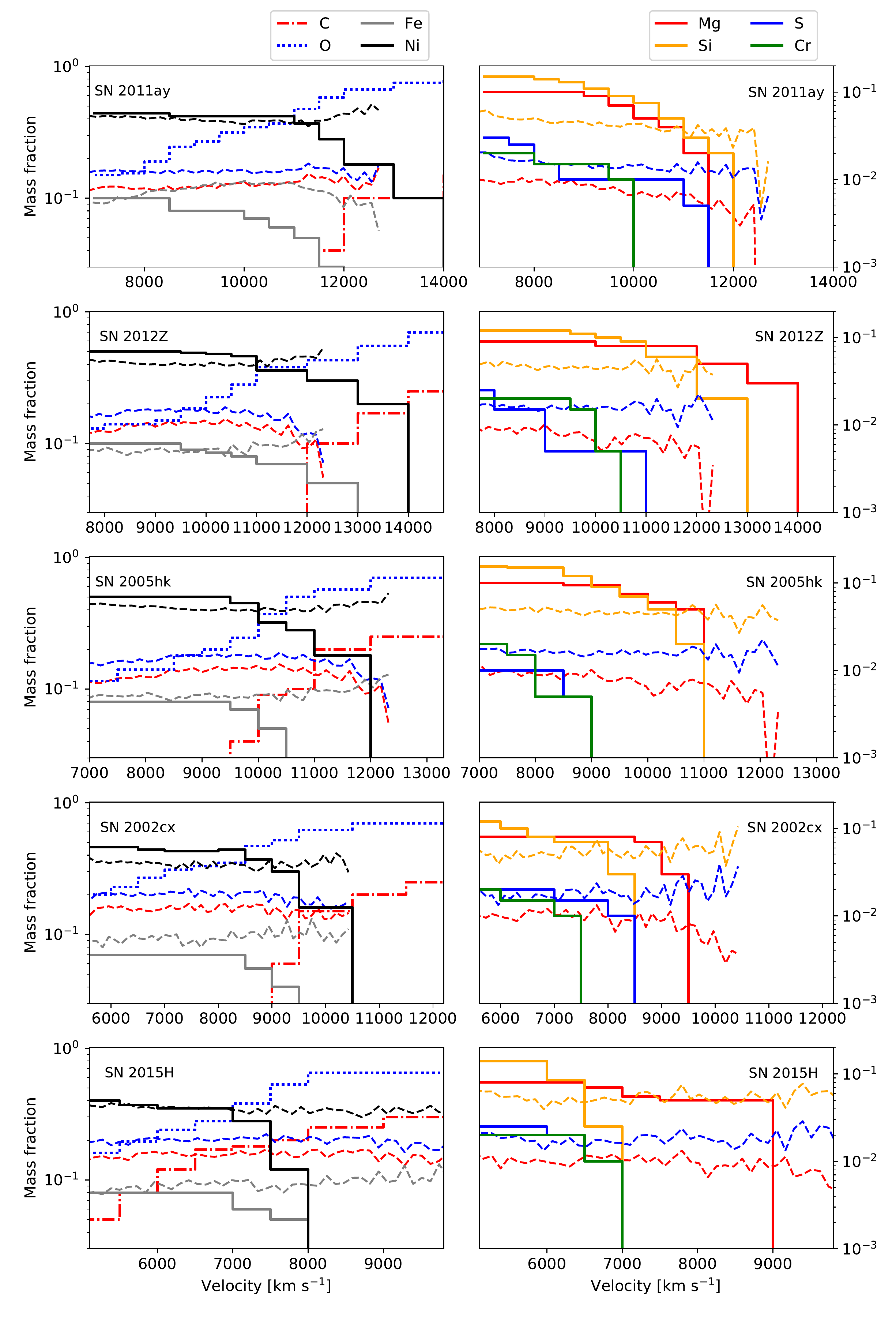}
\caption{\label{fig:ultimate_abundance} The stratified abundance profiles from the TARDIS fitting process of our SNe Iax sample (solid lines), compared to those originating from the corresponding (see in Sec. \ref{sn11ay_results} - \ref{sn2015H_results}) pure deflagration model (dashed lines). Note that we use oxygen (dotted line) as a ``filler'' element, while carbon mass fraction (dash-dot line) shows the limit for the prediction of the deflagration models. The Fe/Co/Ni mass fractions are computed at 100 seconds after the explosion.}
\end{figure*}

\begin{figure*}
\centering
\includegraphics[width=14cm]{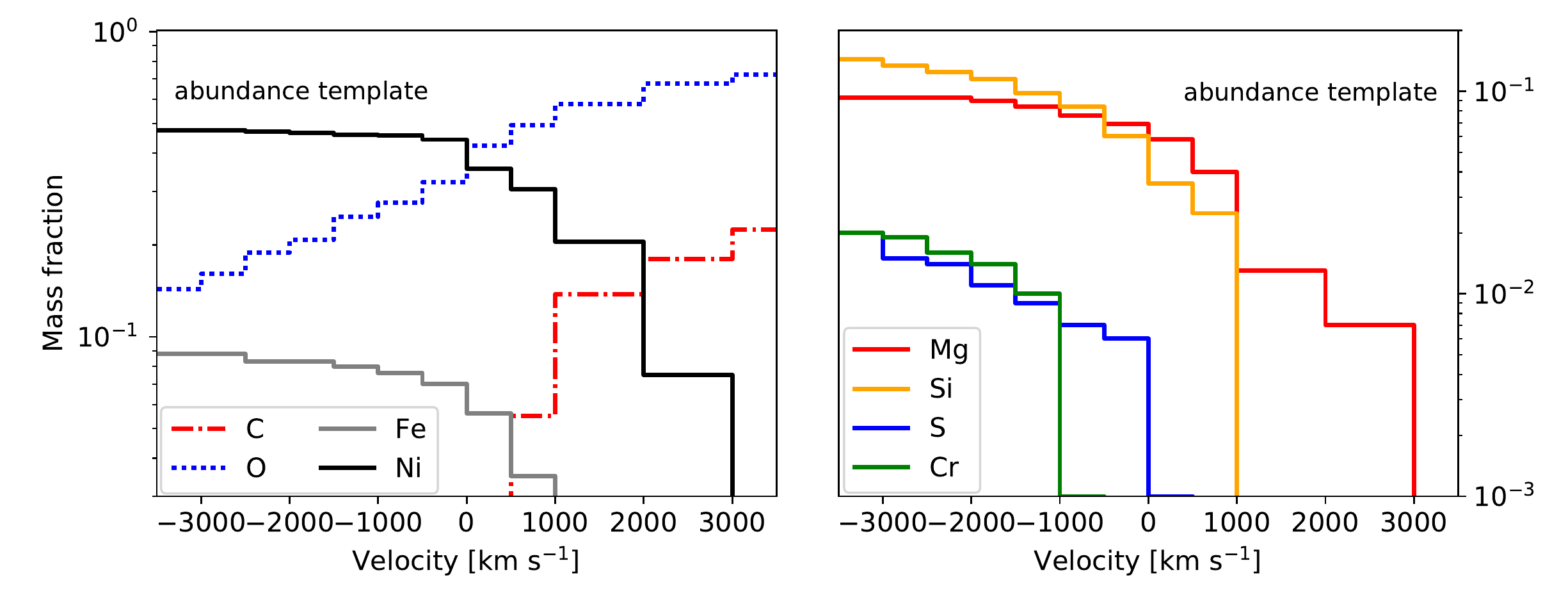}
\caption{\label{fig:template_abundance} The calculated template abundance profile based on the average chemical abundances of the four most luminous SNe in our sample. {The transition velocity, where X(O)=X($^{56}$Ni), is chosen as reference point, thus, it appears at 0 km s$^{-1}$ in this figure}. Note that we use oxygen as a ``filler'' element, while carbon mass fraction shows the limit for the prediction of the deflagration models.The Fe/Co/Ni mass fractions are computed at 100 seconds after the explosion.}
\end{figure*}

%\begin{figure}
%\centering
%\includegraphics[width=\columnwidth]{abundance_sn11ay.pdf}
%\caption{\label{fig:sn11ay_abundance} The stratified abundance profile from the TARDIS fitting process of SN 2011ay spectra (solid lines), compared to those originating from the N10def pure deflagration model (dashed lines). Note that we use oxygen (dotted line) as a ``filler'' element, while carbon mass fraction (dash-dot line) shows the limit for the prediction of the deflagration models.}
%\end{figure}

%\begin{figure}
%\centering
%\includegraphics[width=\columnwidth]{abundance_sn12Z.pdf}
%\caption{\label{fig:sn12Z_abundance} Same as Fig. \ref{fig:sn11ay_abundance} for SN 2012Z and compared to N10def pure deflagration model.}
%\end{figure}

%\begin{figure}
%\centering
%\includegraphics[width=\columnwidth]{abundance_sn05hk.pdf}
%\caption{\label{fig:sn05hk_abundance} Same as Fig. \ref{fig:sn11ay_abundance} for SN 2005hk and compared to N5def pure deflagration model.}
%\end{figure}

%\begin{figure}
%\centering
%\includegraphics[width=\columnwidth]{abundance_sn02cx.pdf}
%\caption{\label{fig:sn02cx_abundance} Same as Fig. \ref{fig:sn11ay_abundance} for SN 2002cx and compared to N3def pure deflagration model.}
%\end{figure}

%\begin{figure}
%\centering
%\includegraphics[width=\columnwidth]{abundance_sn15H.pdf}
%\caption{\label{fig:sn15H_abundance} Same as Fig. \ref{fig:sn11ay_abundance} for SN 2015H and compared to N3def pure deflagration model.}
%\end{figure}

\subsection{Fitting methods}
\label{fitting_method}

In this study, we fit all available spectra of each object with a self-consistent ejecta model, changing only the time-dependent parameters. These are the time since explosion, the inner boundary of the modeling volume, the mass fractions of the radioactive isotopes $^{56}$Ni and $^{56}$Co (which do not result in any additional free parameters in our fitting), and the luminosity, whose change follows the light curve of the SN.

Explosion dates and the derived times since explosions ($t_\rmn{exp}$) of the modelled SNe are from earlier studies (see Table \ref{tab:log}), but we allow them to vary within their uncertainty range, which is usually $\pm$ 1.5 days. The same strategy was used for the emergent luminosity ($L_\rmn{e}$) parameter, which is calculated from the quasi-bolometric light curves of the earlier studies (except for SN 2015H), with an uncertainty range of 0.1 dex. 

The \begin{small}TARDIS\end{small} input parameter called ``velocity of the inner boundary'' is the bottom border of the computation volume, where a blackbody radiation field is emitted. This description shares similarities with the definition of the photosphere, the sharp boundary between the  optically thick and thin parts of the atmosphere. Although a real photosphere cannot be expected in Type Iax SN atmospheres, a mean photosphere is often used in the literature. For these reasons, we call the inner boundary parameter photospheric velocity ($v_\rmn{phot}$) hereafter, which is a free fitting parameter in our method.

The upper end of the computation volume is not strictly limited, but the impact of the outer region rapidly decreases with density. The outer boundary is set sufficiently high 4\,000 km s$^{-1}$ above $v_\rmn{phot}$ for the first spectral fitting of each object and we keep it fixed for the later epochs. The total studied velocity range of an object is higher than 4\,000 km s$^{-1}$, since $v_\rmn{phot}$ decreases with time. Note that the same velocity region, which has negligible impact on the spectrum at later epochs, could be critical at earlier epochs. The model atmosphere is divided into radial velocity shells using steps of 100 km s$^{-1}$. Since we manually fit the spectra, the number of fitting parameters has to be limited. Thus, the mass fractions of the chemical elements are set in a rougher velocity grid with velocity steps of 500 km s$^{-1}$.

{The densities and abundances of our initial models are based on the results of the available hydrodynamical calculations of the pure deflagration scenario, which show the most similarities with the observed properties of SNe Iax. The density and abundance profiles of the model grid presented by \cite{Fink14} are the starting points in our fitting process and we modify the corresponding values if the fitting requires it. In practice, the empirical density profiles used in our simulations are exponential functions:}
\\
\begin{equation}
\rho (v,t_{\rmn{exp}}) = \rho_0 \cdot \left(\frac{t_{\rmn{exp}}}{100\:s}\right)^{-3} \cdot \exp\left({-\frac{v}{2300\:km\,s^{-1}}}\right) \cdot 64^{\frac{v\,-\,v_\rmn{cut}}{v_\rmn{cut}}},
\end{equation}
\\
with a cut off toward higher velocities, which can roughly match the basic structure of deflagration models \citep{Fink14}. These density functions are described by two fitting parameters, the value of the central density, $\rho_{\rmn{0}}$ \citep{Barna17, Magee16} and the location of the cut-off, $v_\rmn{cut}$ (see Table \ref{tab:log_tardis}), where the adopted density profile starts to deviate from the exponential function.

We only use chemical elements that produce unambiguous spectral features for at least one of the observed epochs. These elements are C, O, Na, Mg, Si, S, Ca, Cr, Fe, Co, and Ni for our sample, however, the composition can vary from object to object. Elements with no or weak spectral features are not identified during our analysis, which may result in an unknown mass fraction. Based on the experience of the abundance tomography of SN 2011ay \citep{Barna17}, a significant fraction of mass is not covered by the above mentioned elements, especially in the outermost regions. This could be caused by the presence of unidentified hidden elements, or by an incorrect choice of the density profile. To avoid any false detections of the elements, we choose oxygen as a ``filler'' element to cover the missing mass fraction. Since oxygen forms strong line only at $\lambda$7\,771 \r{A} a week after the explosion, which is close to being saturated, increasing  its fraction in the model atmosphere does not have a significant impact on the spectrum. This also means that the increasing oxygen abundances (see in Sec. \ref{results_outer}) toward higher velocities are probably artificial by-products of our fitting method.

Contrary to our previous work \citep{Barna17}, we use carbon in all of our ejecta models. 
Since carbon is not formed in the SN atmosphere as a nuclear burning product, but comes from the original matter of the progenitor WD, it is a key element in constraining the progenitor and explosion mechanism of a supernova. C\,\textsc{ii} $\lambda$6580 was detected in several SNe Iax \citep{Chornock06,Foley10,McClelland10,Parrent11,Thomas11}, while \citet{Foley13} suggested that every SN Iax may show carbon features in their spectra before or around maximum light. Carbon appears in the theoretical models for SNe Iax with a constant mass fraction of 0.1-0.2 depending on the kinetic energy of the pure deflagration scenario.
However, \cite{Barna17} did not find evidence for the appearance of carbon lines in the spectra of SN 2011ay, thus, carbon was not included in that study. To revise the results of \cite{Barna17} and extend our investigation of the unburnt material, we have changed our fitting strategy in relation to carbon. Here, we try to discover the upper limit of the element both in location and mass fraction.

To summarize the free parameters of the individual model of each SN are time of explosion, the central density $\rho_{\rmn{0}}$ and the location of the density cut-off $v_\rmn{cut}$. The latter two define the density profile. The mass fractions of the chemical elements in each radial shell are also free parameters. Using the same ejecta model for fitting all the spectra of the same SN keeps the self-consistency of our modeling approach. The $L_\rmn{e}$ and $v_\rmn{phot}$ are fit for all the available spectra. We explore the parameter space until we find a good visual agreement between the observations and the synthetic spectra.

\section{Results and discussion}
\label{results}

Figs. \ref{fig:sn11ay_spectra} - \ref{fig:sn15H_spectra} show that our fitted density and abundance profiles are able to reproduce the flux continuum and the main spectral features in almost all cases.

We list the fitted physical parameters ($t_\rmn{exp}$, $L_\rmn{e}$, $v_\rmn{phot}$) and the density profile defining parameters ($\rho_\rmn{0}$ and $v_\rmn{cut}$) for each epoch of the SNe in our sample in Table \ref{tab:log_tardis}. 
The fitted density profiles in our individual models show a somewhat steeper function than in the cases of the corresponding deflagration models (see in Table \ref{tab:log_tardis}). The discrepancies are especially high in the cases of SNe 2002ck, 2005hk and 2012Z, where the fitting of the steep flux continuum and of the blue wings of spectral lines requires a stronger cut-off at the higher velocities in the ejecta. In order to illustrate the impact of the cut-off, we compare the effect of the adopted density profile for SN 2012Z with the density function of the N5def model and the same model without a cut-off at the higher velocities. As it can be seen in Fig. \ref{fig:density_sn12Z}, a stronger cut-off (lower $v_\rmn{cut}$ value) in the density profile reduces the excessive blueshifts of the spectral lines of IGEs.

All the five best-fit models show stratified chemical structures (see Fig. \ref{fig:ultimate_abundance}), where the mass fractions of almost all the chemical elements change with velocity (for the parameters of the individual abundance profiles, see Tables \ref{tab:data_sn11ay} - \ref{tab:data_sn15H}). These abundance profiles show many similarities in both density and abundance profiles despite being fit for different supernovae.  Note that these findings are in tension with the deflagration models which are taken as initial point for our fitting process (see Sec. \ref{fitting_method}), since those hydrodynamical calculations predict nearly constant mass fractions for each element. Based on the same features in the abundance profiles, we create an abundance template for our sample.

In the following, we use the abundance profiles to create a few-parameter description of SN Iax, which we call the template abundance profile (Sec. \ref{results_template}). We compare this template to the general predictions of the pure deflagration models and use it to describe the above mentioned similar structures of the abundance profiles(see Sec. \ref{results_inner} and \ref{results_outer}). The sections \ref{sn11ay_results} - \ref{sn2015H_results} describe the unique attributes of each best-fit model and, if possible, we compare it to other spectroscopic analyses in the literature.

\subsection{Abundance template for the Type Iax sample}
\label{results_template}

We calculate the mean chemical abundances of the SNe (except SN 2015H, see Sec. \ref{sn2015H_results}) in each radial shell to create an abundance template model for the whole Type Iax sample. For this purpose, we choose the transition velocity in every abundance profile, where the mass fraction of oxygen is equal with that of radioactive nickel (X(O) = X($^{56}$Ni)), as a reference point. The fitted individual abundance profiles (Fig. \ref{fig:ultimate_abundance}) are shifted in velocity space to match the same transition velocity. Then, we calculate the average mass fractions in each radial shell. The resulting abundance profiles of the chemical elements can be seen in Fig. \ref{fig:template_abundance}.

The template abundance profile preserves the common characteristics of the individual objects, like the decreasing abundance of both IMEs and IGEs with increasing velocity or the lack of carbon in the inner regions. However, these features exist at different velocities from object to object in our sample, roughly corresponding to their $v_\rmn{phot}$values at maximum light. {To test our hypothesis, we use the abundance template (Fig. \ref{fig:template_abundance}) with transition velocities of 7\,000, 8\,500, 10\,000 and 11\,000 km s$^{-1}$ for SNe 2015H, 2002cx, 2005hk and 2011ay, respectively, according to the transition velocity of our fits of each object (see Sec. \ref{sn11ay_results} - \ref{sn2015H_results}).} We generate synthetic spectra for one epoch of each object and compare them to both the observed spectra and our best-fit models. To demonstrate the effect of the same abundance profile at completely different velocities, we choose similar epochs in pairs: approximately eight days after the explosion of SN 2002cx and SN 2005hk, and 20 days in the case for SN 2015H and SN 2011ay. Since our goal is only to test the uniformity of the abundance profile, the luminosities and the times of the explosions, just like the density profiles, remain the same as in the best-fit models.

\begin{figure}
\centering
\includegraphics[width=\columnwidth]{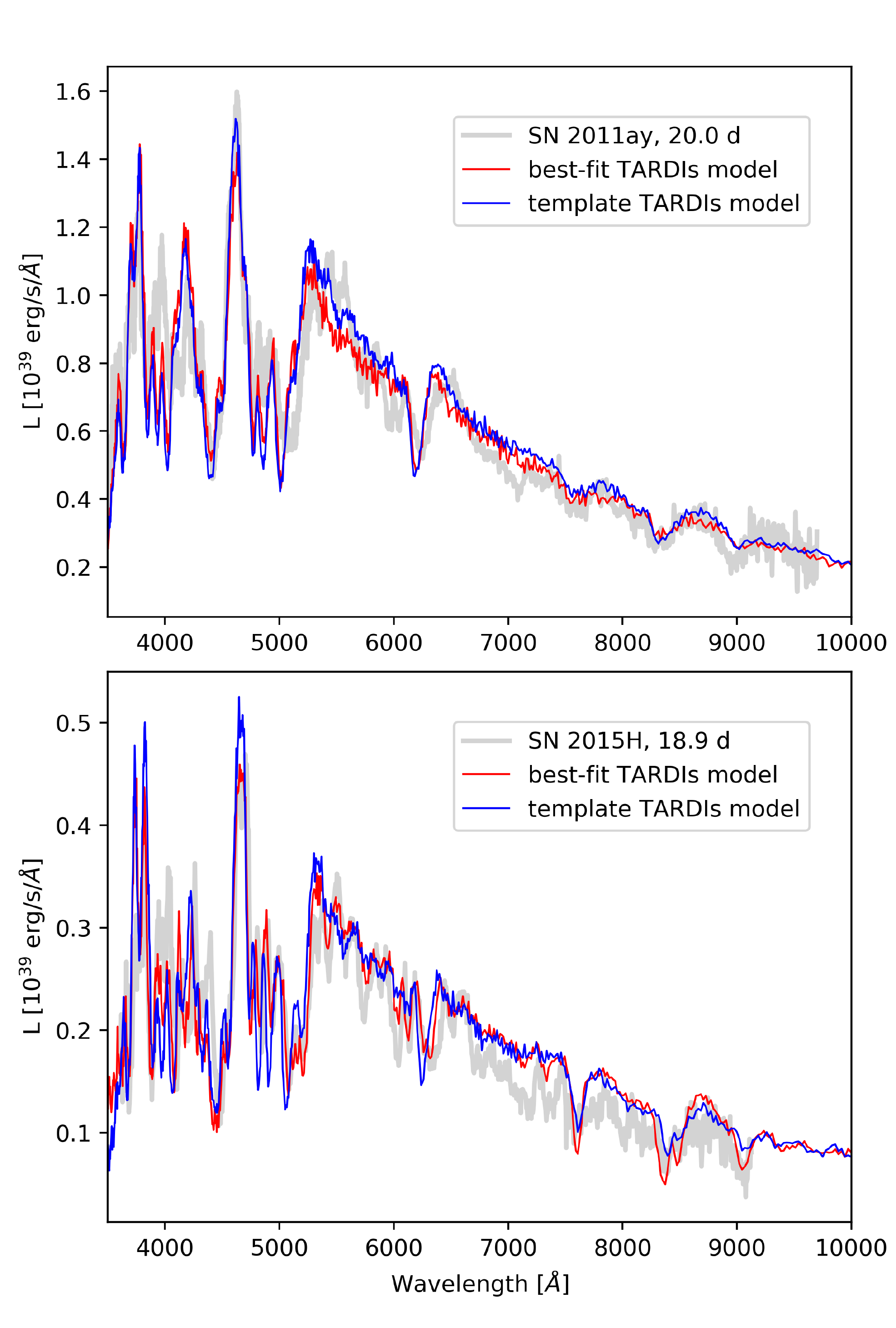}
\caption{\label{fig:template1} Spectra of SN 2011ay and SN 2015H obtained at nearly the same epoch (grey), compared to the synthetic spectra calculated with the template abundance profile (blue). The transition velocity of the template (Fig. \ref{fig:template_abundance}) is shifted to 7\,000 and 11\,000 km s$^{-1}$ for SN 2015H and SN 2011ay, respectively.}
\end{figure}

\begin{figure}
\centering
\includegraphics[width=\columnwidth]{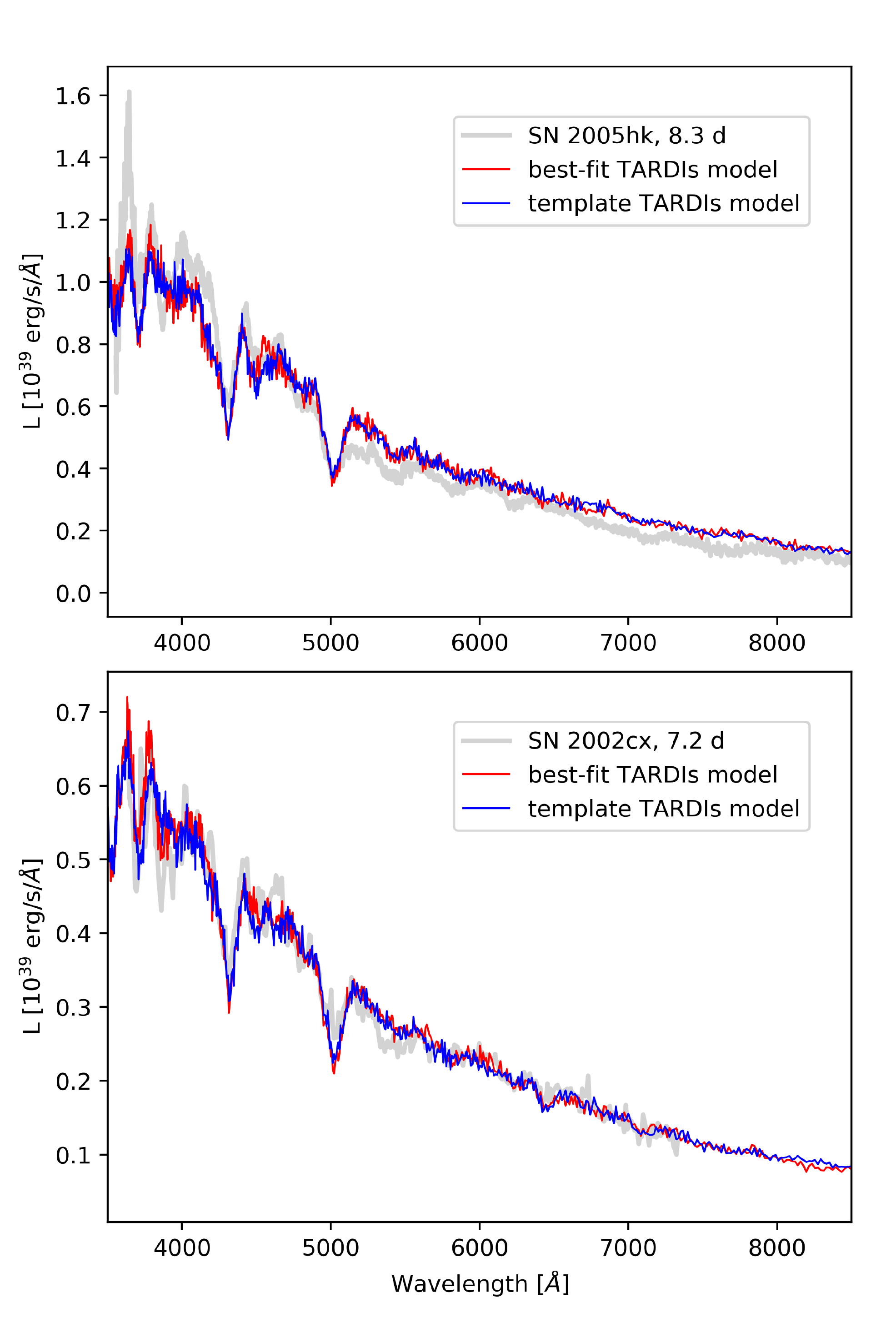}
\caption{\label{fig:template2} Spectra of SN 2005hk and SN 2002cx obtained at nearly the same epoch (grey), compared to the synthetic spectra calculated with the template abundance profile (blue). The transition velocity of the template (Fig. \ref{fig:template_abundance}) is shifted with 8\,500 and 10\,000 km s$^{-1}$ for SN 2002cx and SN 2005hk, respectively.}
\end{figure}

The effect of the template profile can be seen in Figs. \ref{fig:template1} and \ref{fig:template2}, where the quality of the fits weakens only slightly in general. This indicates that we could effectively describe the chemical structure of the SNe in our sample with only one parameter. The fact that the same template abundance profile describes SNe with completely different luminosities increases the possibility that Type Iax SNe have the same or a similar progenitor scenario. The used deflagration-like density profiles defined only with two free parameters (see Sec. \ref{fitting_method}) also support this assumption. However, at this point, we cannot state that the whole subclass is a one-parameter family like the normal Type Ia SNe, but it seems to be possible that only a few free parameters (velocity shift of the abundance profile, density function) can efficiently describe all the SNe Iax.

\subsection{Inner regions of the abundance template}
\label{results_inner}

Figure \ref{fig:template_abundance} shows that the inner layers of the template model atmosphere are dominated by IGEs. The most abundant species is the radioactive $^{56}$Ni that reaches up to a fraction of 0.40-0.50 in the innermost ejecta and gradually decreases toward the outer regions. The extension of $^{56}$Ni and its decay products in our best fit models is limited by the blue side of their absorption lines at post-maximum epochs. The initial mass fractions of stable iron roughly follows the changes of $^{56}$Ni, varying between 0.05 and 0.15. The amount and location of stable iron is constrained by the strong Fe\,\begin{small}III\end{small} lines at $\sim$4400 and 5600 \r{A} in the pre-maximum spectra.

IMEs appear with smaller mass fractions than IGEs at the same velocities. The mass fractions of Si and S are well constrained, because of the relatively strong, but not saturated Si\,\begin{small}II\end{small} $\lambda$6\,355, Si\,\begin{small}III\end{small} $\lambda$4\,588 lines and the S\,\begin{small}II\end{small} ``W'' absorption feature. The mass fractions of these two elements follow a similar trend between 0.01 and 0.10, but silicon is always more abundant than sulfur. Magnesium appears with mass fractions of 0.05-0.10 with higher uncertainty, but it also disappears at the highest velocities just like other IMEs. Calcium abundances, which are also well constrained based on the fitting of the Ca\,\begin{small}II\end{small} H$\&$K lines and the Ca II NIR triplet, stay below 0.01 in all the cases.

Note that despite using a stratified abundance structure, our template estimated from the best-fit models does not differ strongly in the inner regions from the prediction of the pure deflagration models. Indeed, the amount of most elements (O, Si, S, Ca, Fe, $^{56}$Ni), under a certain velocity, show similar mass fractions to those of the corresponding pure deflagration models of \cite{Fink14}. 

However, a few discrepancies also appear in these comparisons. In the case of magnesium, one order of magnitude higher mass fraction is necessary for our \begin{small}TARDIS\end{small} model to fit the absorption feature of Mg\,\begin{small}II\end{small} at 9500 \r{A} at least partially. Chromium and titanium show only a modest fraction in our best-fit models, although, even these values exceed the predictions of the deflagration model by one and three orders of magnitude, respectively. The amount and the location of chromium is well-constrained in our analysis, thanks to the several spectral features of Cr\,\begin{small}II\end{small} ($\lambda$4\,588, $\lambda$5\,237 and the flux suppression at $\sim$3\,800 \r{A}). 

Titanium is only required to fit the blue wing of the absorption feature at $\sim$4\,300 \r{A}. The Ti\,\begin{small}II\end{small} features in this wavelength region are strongly temperature-dependent \citep{Hatano99}. Although, the radiation temperatures derived from the \begin{small}TARDIS\end{small} simulations are robust (see Sec. \ref{method}), the assumptions of the level populations, for which radiation temperature is the dominating parameter, are not \cite[for further description see ][]{Kerzendorf14}. Thus, we cannot strongly constrain the amount of titanium based only on this feature.

Carbon is not allowed in the inner regions in our template model. This finding is a further discrepancy to the hydrodynamical calculations of the deflagration scenario, since carbon is the third most abundant element at any velocity in those models. Significant amounts of carbon cause strong characteristic spectral lines of C\,\begin{small}II\end{small} $\lambda$4\,764$,  \lambda$6\,578 as can be seen in Figs. \ref{fig:sn11ay_spectrum_ions} - \ref{fig:sn02cx_spectrum_ions}.

\subsection{Outer regions of the abundance template}
\label{results_outer}

Figure \ref{fig:template_abundance} shows that the outer regions of the template model are dominated by oxygen reaching mass fractions of 0.80. However, since we use oxygen as a ``filler'' element (see Sec. \ref{fitting_method}), this feature might be artificial. Thus, we cannot draw any conclusion about the overabundance of oxygen. 

Apart from oxygen, carbon and sodium are the only elements, whose appearance is allowed in the outermost layers of the model ejecta. We increased the mass fraction of Na up to 0.05 in the colder outer layers in order to reproduce the characteristic Na\,\begin{small}I\end{small} feature $\lambda$5\,890 at the later epochs. Note that this amount of Na is extreme compared to the deflagration models, which show two orders of magnitude lower values.

\begin{figure*}
\centering
\includegraphics[width=130mm]{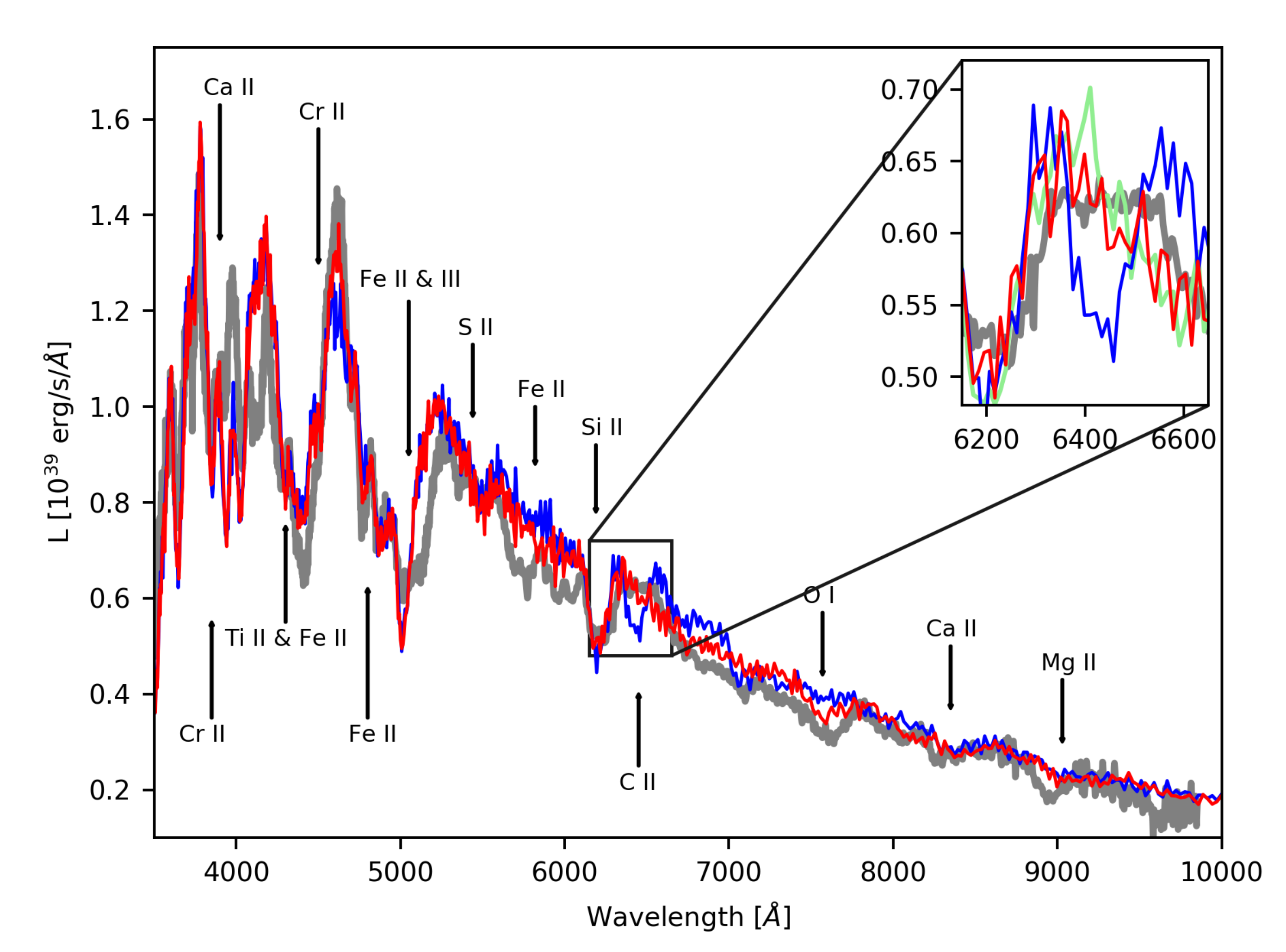}
\caption{\label{fig:sn11ay_spectrum_ions} The spectrum of SN 2011ay at the epoch +2.6 days with respect to B-maximum (grey), the synthetic spectrum from the best-fit model (red), the same model with constant (blue) carbon mass fraction of X(C)=0.12 and with zero carbon mass fraction (green, only in the cut-out). The black arrows show the position of prominent absorption features and the related ions.}
\end{figure*}

\begin{figure*}
\centering
\includegraphics[width=130mm]{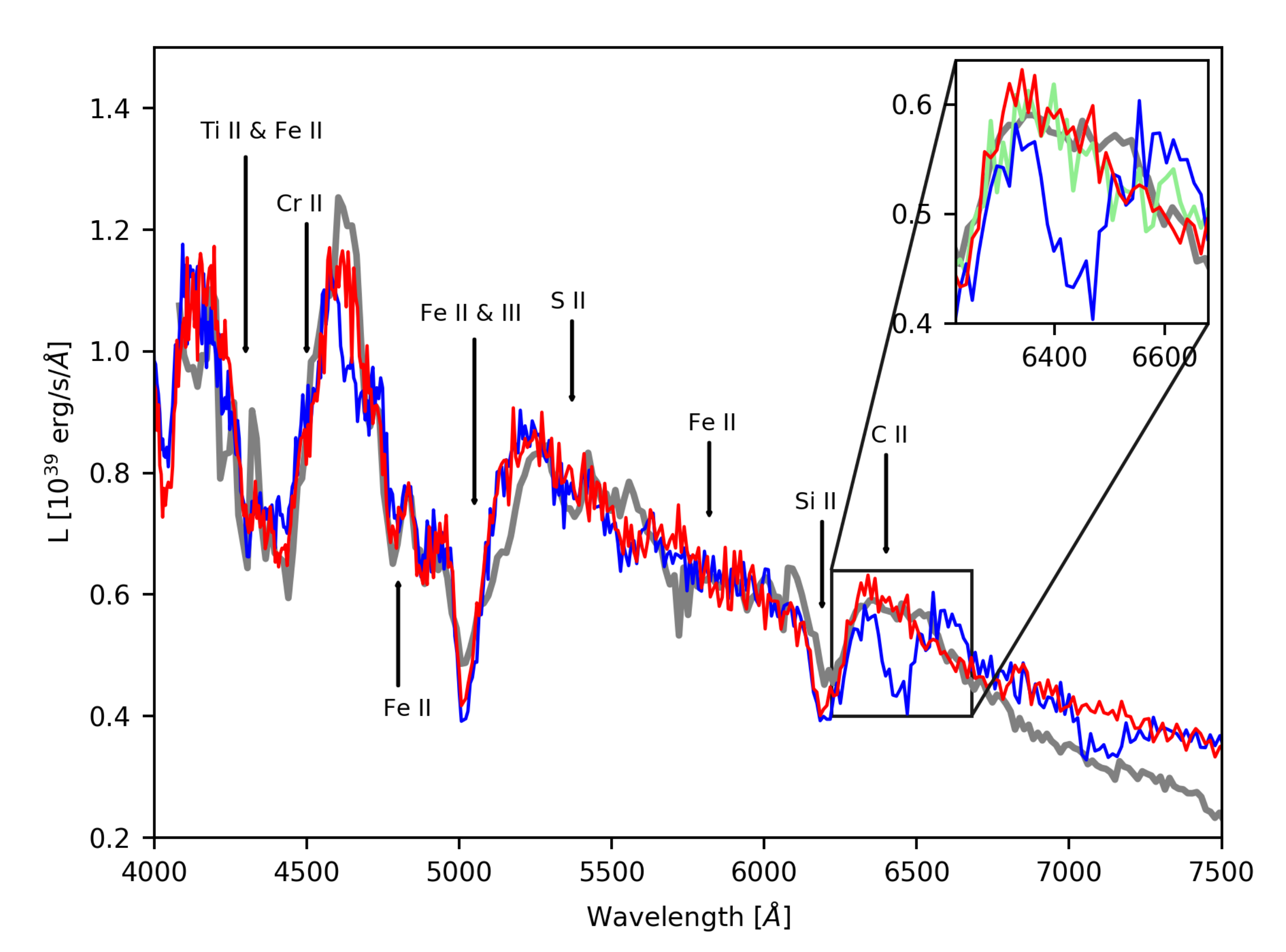}
\caption{\label{fig:sn12Z_spectrum_ions} Same as Fig. \ref{fig:sn11ay_spectrum_ions} for SN 2012Z at the epoch +1.1 days with respect to B-maximum.}
\end{figure*}

Although carbon does not appear in the inner regions of the template model, its presence is allowed in the outer layers with nearly the same mass fraction as in the deflagration models. However, the complete elimination of carbon changes the goodness of the fits only slightly. To highlight the constraints on carbon in our best-fit models, we also plot spectra estimated from models with constant carbon fraction, predicted by the deflagration models, and from models with zero carbon, in Figs. \ref{fig:sn11ay_spectrum_ions} - \ref{fig:sn15H_spectrum_ions}.

\subsection{SN 2011ay}
\label{sn11ay_results}

\cite{Barna17} fixed the density function of the model of SN 2011ay to the exponential fit of the W7 density profile \citep{Nomoto84}. Since we now try to connect our \begin{small}TARDIS\end{small} models to hydrodynamical calculations, the densities are fitting parameters within the frame of the above described deflagration-like profiles. As a result, the new density function of the SN 2011ay model is similar to that of the N10def deflagration model (see Table \ref{tab:log_tardis}), whose peak luminosity ($M_\rmn{V}$ = -18.38 mag) is the closest to SN 2011ay. 

%Integrating the assumed density profile, the total ejecta mass in our model is M$_{ej}$ = 0.64 $M_{\odot}$, which is slightly lower than the M$_{ej}$ = 0.86 $M_{\odot}$ calculated from the N10def model.\footnote{Note that the total ejecta mass of our model comes from the integration of the whole density function, not just from the velocity frame covered by our modeling process. Such extrapolation is reasonable, since both the deflagration models and our findings show an exponential structure for the innermost regions.}

The abundance profiles are also subject to revision. However, despite the modification of the ejecta profile and the appearance of new elements (sodium and carbon) in our model, the primary attributes of the chemical abundances remain the same and the conclusions stated in \cite{Barna17} still hold. The adopted time of the explosion from the recent fitting process is $T_\rmn{0}$ = 55632.7 MJD, which is 0.2 days later than in our previous paper. The changes of the density profile affect mainly the longer wavelengths, at pre-maximum epochs, where the flux level of the synthetic spectra gets closer to the observations (see Fig. \ref{fig:sn11ay_spectra}). This is a significant improvement in the quality of the fits. However, around and after maximum light, the blue wings of the strongest absorption lines are less well fit, because of the lack of material at higher velocities. On the whole, the quality of the fits does not change significantly.

The abundance profile coming from our best-fit \begin{small}TARDIS\end{small} model can bee seen in Fig. \ref{fig:ultimate_abundance}. Comparing these mass fractions to those calculated from the N10def pure deflagration model, we find that the amount of most elements (O, Si, S, Ca, Fe, $^{56}$Ni) show similar mass fractions as the deflagration model. The total mass of the radioactive nickel predicted by the N10def model (0.26 $M_{\odot}$) is also close to the estimated value from the quasi-bolometric light curve \citep[0.22 $\pm$ 0.01 $M_{\odot}$; ][]{Szalai15}. 

The abundances start to differ strongly only above the transition velocity, $\sim$11\,000 km s$^{-1}$, where the $^{56}$Ni abundance drops below the increasing mass fraction of oxygen. A further discrepancy between our findings and the N10def model is the appearance of carbon (see Fig. \ref{fig:sn11ay_spectrum_ions}), which is strictly constrained to the outer part of the ejecta in our analysis. However, above 12\,000 km s$^{-1}$, even the carbon mass fraction (0.10-0.15) meets the prediction of the deflagration model.

\subsection{SN 2012Z}
\label{sn12Z_results}

As can be seen in Fig. \ref{fig:sn12Z_spectra}, the model fits for this supernova are not as good as for SN 2011ay in general.  Although the primary features can be found in the synthetic spectra (see Fig. \ref{fig:sn12Z_spectrum_ions}), several IGE lines are poorly fit before (especially iron) and after (cobalt and nickel) maximum. The continuum is well reproduced, but strong deviations appear at the near-infrared end of the spectrum near maximum light. 

We found the date of the explosion $T_\rmn{0}$ = 55952.8 MJD. Comparing the emergent luminosity parameter to that of SN 2011ay, SN 2012Z seems slightly fainter at the same epochs. This contradicts the result of \cite{Stritzinger14}, but such moderate discrepancy could arise from the uncertainty of the distance.

The density profile of the model of SN 2012Z is between those of the N5def and N10def deflagration models, although, with a steeper cut-off at higher velocities (see Table \ref{tab:log_tardis}). This feature of the density profile is necessary to reproduce the blue wings of the spectral lines (Fig. \ref{fig:density_sn12Z}). 
Since the photometry of SN 2012Z is better matched by the N10def model (peak absolute magnitude of $M_\rmn{V}$ = -18.38 mag), the abundance profile is compared to its predictions in Fig. \ref{fig:ultimate_abundance}. Like for SN 2011ay, we find that the inner part of the ejecta shares similarities to the predictions of the deflagration model. The transition velocity between the $^{56}$Ni dominated inner- and the O dominated outer part of the model ejecta is 11\,000 km s$^{-1}$. Carbon appears only above this velocity limit in our model. 

%The total ejecta masses of the adopted density model and N10def are very similar (M$_{ej}$ = 0.49 and 0.48 $M_{\odot}$, respectively).

%{SN 2012Z is to date the only thermonuclear SN with a detected candidate companion star.} The assumption of a He-donor companion star does not create robust constraints in the observables, although the appearance of helium is expected at least in the outer regions of the atmosphere. Since He\,\begin{small}I\end{small} is expected to form only one optically thin spectral line ($\lambda$10830 \r{A}) in the optical/near-infrared regime in the temperature range of a typical SN \citep{Boyle17}, which is, moreover, blended with Fe\,\begin{small}II\end{small} and Mg\,\begin{small}II\end{small} lines, its identification would be highly ambiguous. According to the description of our method above, we had eliminated helium from our modeling process.

\begin{figure*}
\centering
\includegraphics[width=130mm]{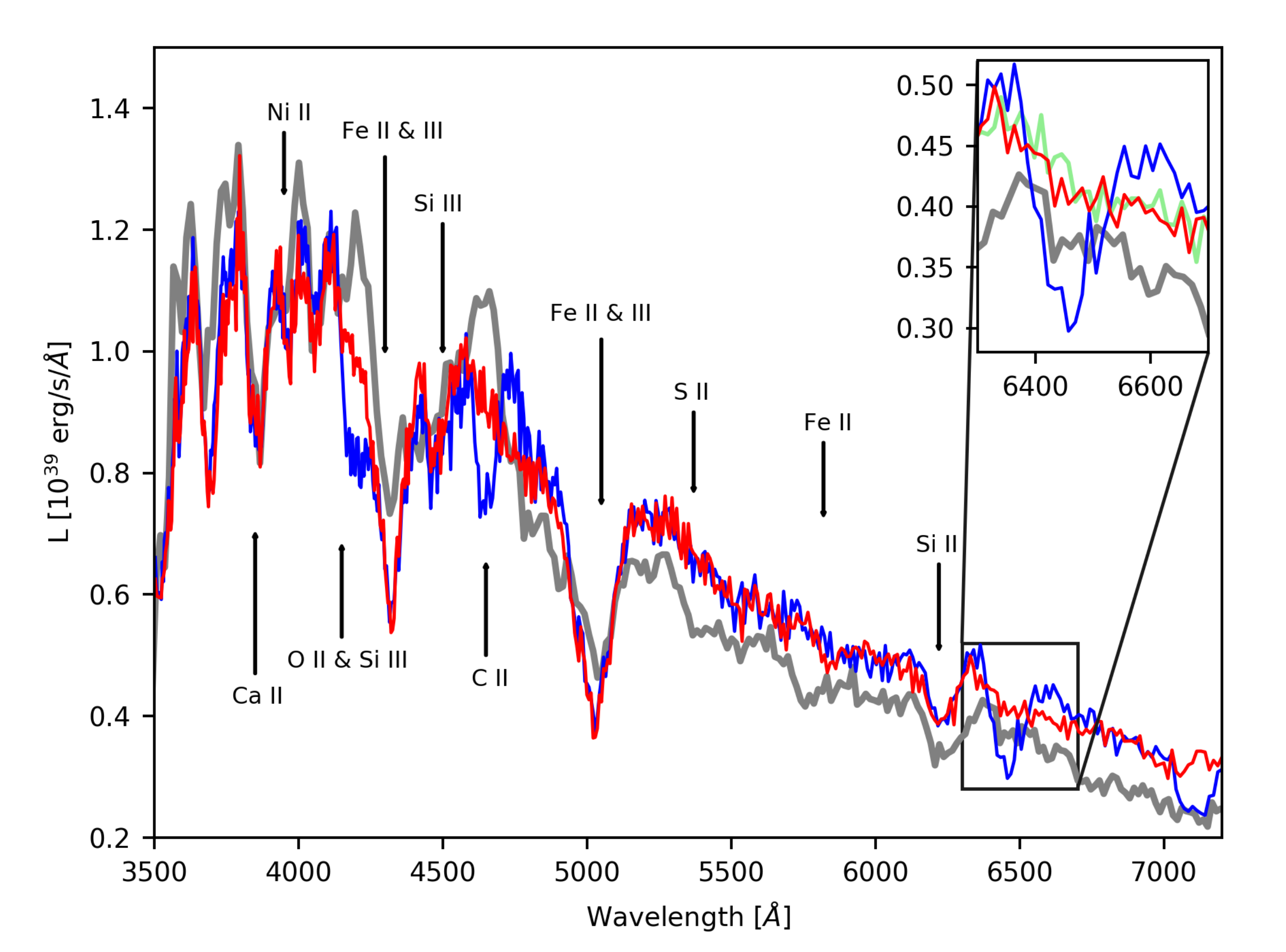}
\caption{\label{fig:sn05hk_spectrum_ions} Same as Fig. \ref{fig:sn11ay_spectrum_ions} for SN 2005hk at the epoch -1.3 days with respect to B-maximum.}
\end{figure*}

\begin{figure*}
\centering
\includegraphics[width=130mm]{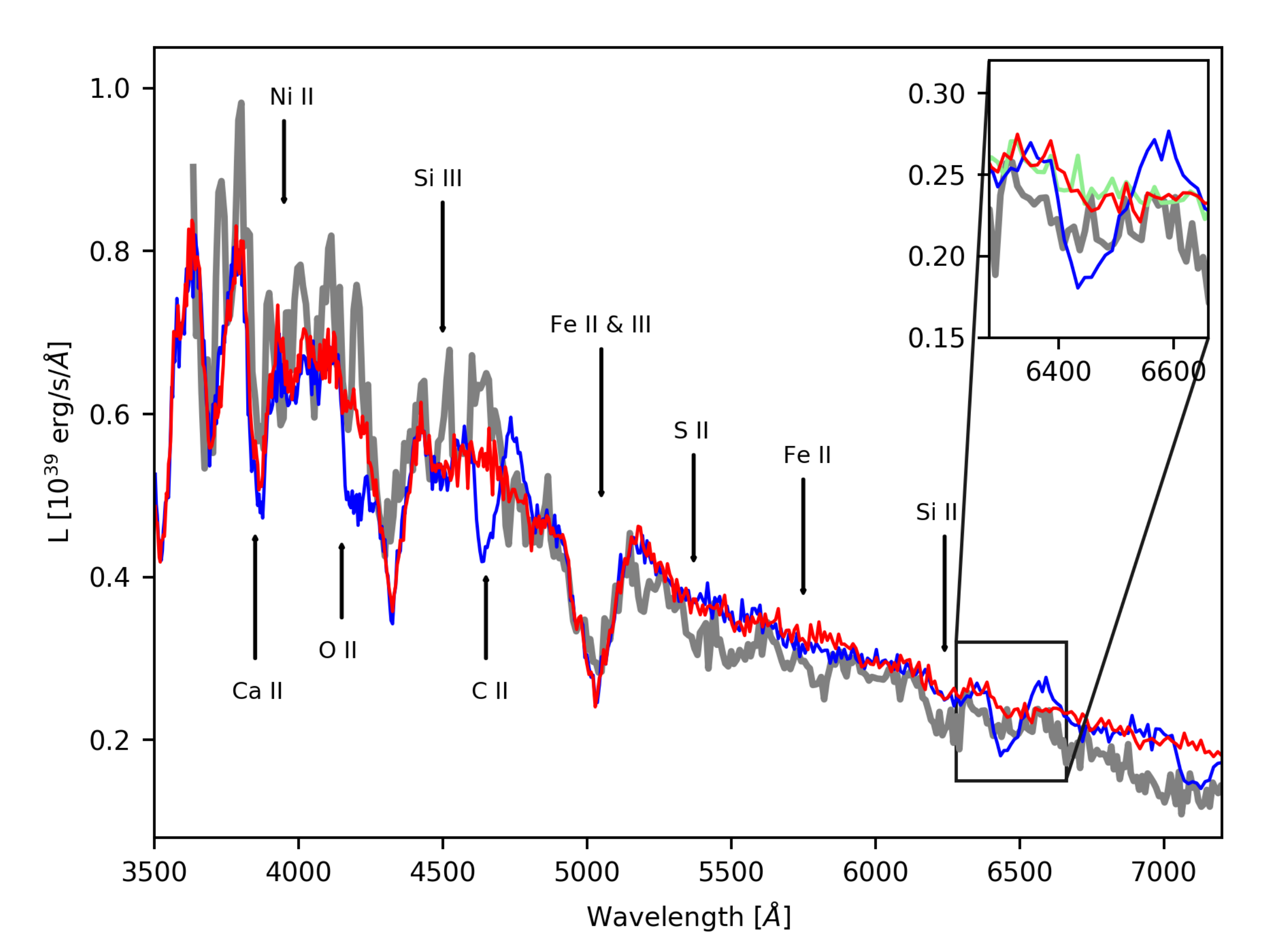}
\caption{\label{fig:sn02cx_spectrum_ions} Same as Fig. \ref{fig:sn11ay_spectrum_ions} for SN 2002cx at the epoch -1.0 days with respect to B-maximum.}
\end{figure*}

\begin{figure*}
\centering
\includegraphics[width=130mm]{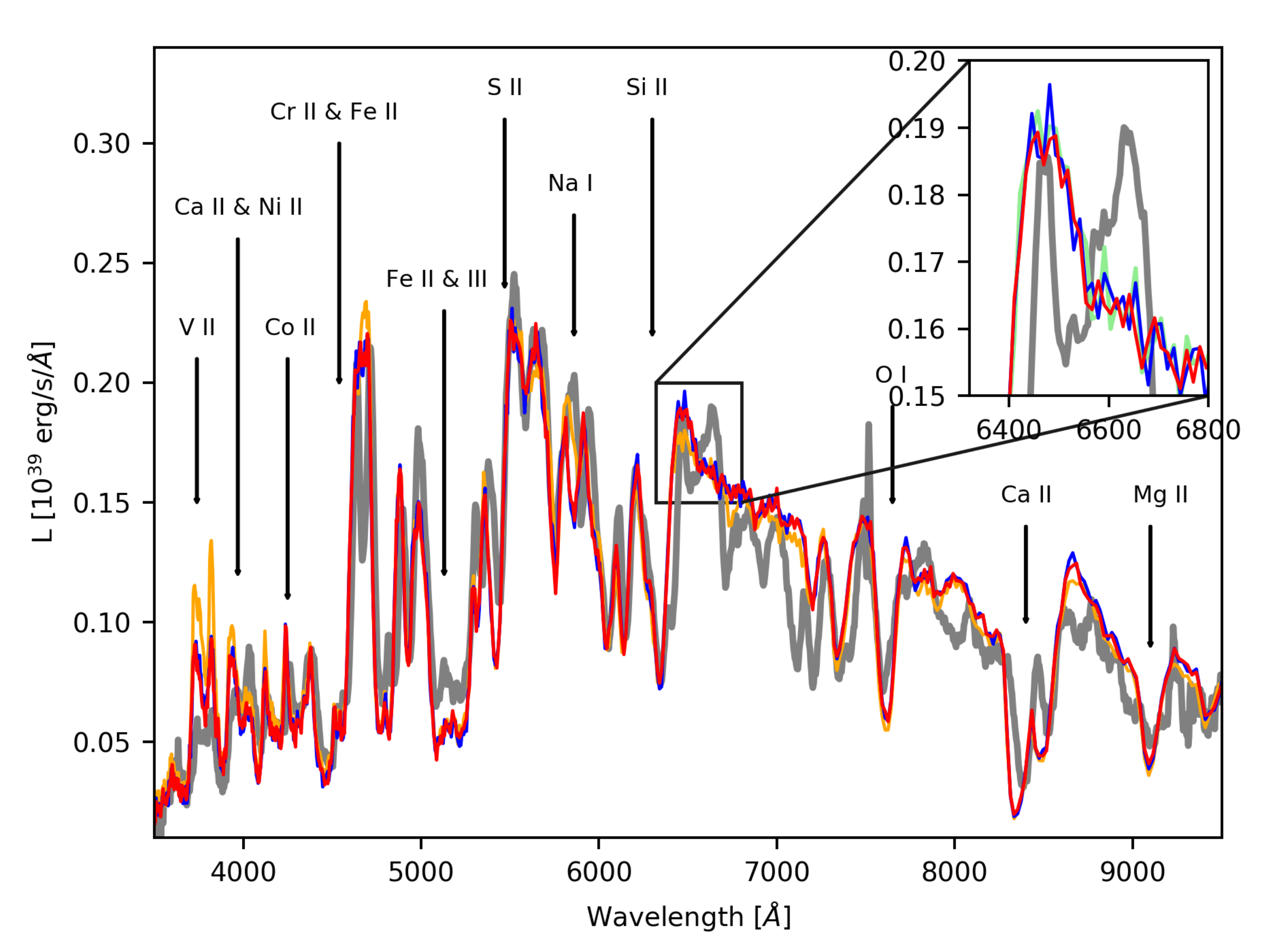}
\caption{\label{fig:sn15H_spectrum_ions} Same as Fig. \ref{fig:sn11ay_spectrum_ions} for SN 2015H at the epoch +10.4 days with respect to r-maximum. The vanadium-free TARDIS models is also plotted (orange).}
\end{figure*}

\subsection{SN 2005hk}
\label{sn05hk_results}

In case of SN 2005hk, which is the second best observed SN in our sample, the deviation between the observed spectra and our best-fit models are significant (see Fig. \ref{fig:sn05hk_spectra}). This discrepancy occurs especially a few days before maximum light, when the continuum of our synthetic spectrum above 5\,500 \r{A} is consistently too high. The strongest IGE lines at shorter wavelengths are also not fit well. Fixing these mismatches would require an extremely steep density profile, which would differ significantly from the adopted deflagration-like ejecta profiles, and the changes would spoil the relatively good fit at the earliest and latest epochs (see Fig. \ref{fig:sn05hk_spectrum_ions}).

In Fig. \ref{fig:ultimate_abundance}, we show the abundance profile of our fits with the N5def model, whose peak absolute magnitudes ($M_\rmn{B}$ = -17.85 mag and $M_\rmn{V}$ = -18.24 mag) show the best agreement with SN 2005hk ($M_\rmn{B}$ = -18.02 mag and $M_\rmn{V}$ = -18.08 mag).
The transition velocity appears at 10\,000 km s$^{-1}$; lower, than in the case of the more luminous SNe 2011ay and 2012Z.

%, whose ejecta mass (M$_{ej}$ = 0.37 $M_{\odot}$) shows an excellent match with extrapolation of our results to the lower velocities (0.32 $M_{\odot}$). 53675.2 & 5.3 

\cite{Magee17} used \begin{small}TARDIS\end{small} to compare the abundance-profile from the N5def model to the spectrum obtained at +3.6 days after B-maximum. The authors used $T_\rmn{0}$ = 53670.0 MJD as the time of explosion, which agrees with our value.
They found that a deflagration model with a relatively restricted modeling volume between 7\,800 and 9\,400 km s$^{-1}$ can broadly fit the flux continuum and the main features of IGEs and IMEs. However, they also showed that reducing the carbon mass fraction by an order of magnitude over the whole model volume improves the match of the C\,\begin{small}II\end{small} $\lambda$4750 and $\lambda$6580 features. This result is consistent with our best-fit model for SN 2005hk, in which the abundances in the region between 7\,800 and 9\,400 km s$^{-1}$ are very similar to those of the N5def; except for the case of carbon, which is nearly absent from this part of the ejecta in our model.

Beside SN 2011ay, SN 2005hk is the only other SN Iax that has been the subject of a previously published abundance tomography analysis \citep{Sahu08}. Thus, SN 2005hk offers a unique opportunity to compare our \begin{small}TARDIS\end{small} model to the results of an independent investigation. However, it must be borne in mind that \cite{Sahu08} used different radiation-matter approximations, which may cause systematic differences in the modeling process.

\cite{Sahu08} adopted a much steeper density profile for their model ejecta, resulting in significantly more mass at lower velocities. Their densities are near or above the W7 model at the studied epochs, and significantly higher values of the density function than used here. Lower $v_\rmn{phot}$ values were chosen in their fitting process than in our modeling. Despite the different modeling approach and physical parameters, the abundances show a surprisingly good agreement with our results. The earliest model atmosphere, which maps the highest velocity regions, shows an extreme oxygen mass fraction of 0.86. Oxygen remains dominant in the later epochs after maximum light, while the amount of IGEs increases with time. The mass fractions of IMEs are the same order of magnitude like in our \begin{small}TARDIS\end{small} model, while the mass fraction of carbon is under 0.01. All these similarities support the results of our modeling.

\subsection{SN 2002cx}
\label{sn02cx_results}

Only three spectra are available for SN 2002cx within the time range applicable for \begin{small}TARDIS\end{small} analysis. Two of these were obtained before the  B-band maximum, while the last was taken 12 days after maximum. This observational gap means a $\sim$2\,000 km s$^{-1}$ jump between the $v_\rmn{phot}$ values of the corresponding epochs, causing a significant uncertainty in our modeling process. All spectra have low signal-to-noise ratios making the accurate fitting of optically thin lines more difficult. Despite these complicating factors, the quality of the fit is high regarding both the continuum level and the line profiles (see Fig. \ref{fig:sn02cx_spectrum_ions}).

%The adopted density profile for SN 2002cx is between N3def and N5def, but the integrated total ejecta mass, M$_{ej}$ = 0.19 $M_{\odot}$ is closer to the less energetic deflagration model N3def (M$_{ej}$ = 0.20 $M_{\odot}$). 
The abundance profile of our model can be seen in Fig. \ref{fig:ultimate_abundance}. The time of explosion is found as $T_\rmn{0}$ = 52404.0 MJD. The adopted density profile for SN 2002cx is between N3def and N5def, but the absolute peak magnitude is closer to the less luminous deflagration model N3def ($M_\rmn{V}$ = -17.52 mag).

The transition velocity mentioned above, where our models start to deviate from the predictions of hydrodynamic calculations, appears at 8\,500 km s$^{-1}$, i.e. 1\,500 km s$^{-1}$ lower than for SN 2005hk. At the same time, the fitting of the C\,\begin{small}II\end{small} line at 6\,500 \r{A} gives strong constraints on the location of carbon above 9\,000 km s$^{-1}$ as well. Vanadium is used as an additional element to improve the fitting of the latest spectrum at $\sim$3\,800 \r{A}, where V\,\begin{small}II\end{small} suppresses the flux. Note that this is the only strong feature which supports the detection of V. Despite its small derived mass fraction (0.01-0.02), the V abundance in the inner part of the ejecta exceeds the prediction of the deflagration models by a few orders of magnitude.

\subsection{SN 2015H}
\label{sn2015H_results}

\begin{figure}
\centering
\includegraphics[width=\columnwidth]{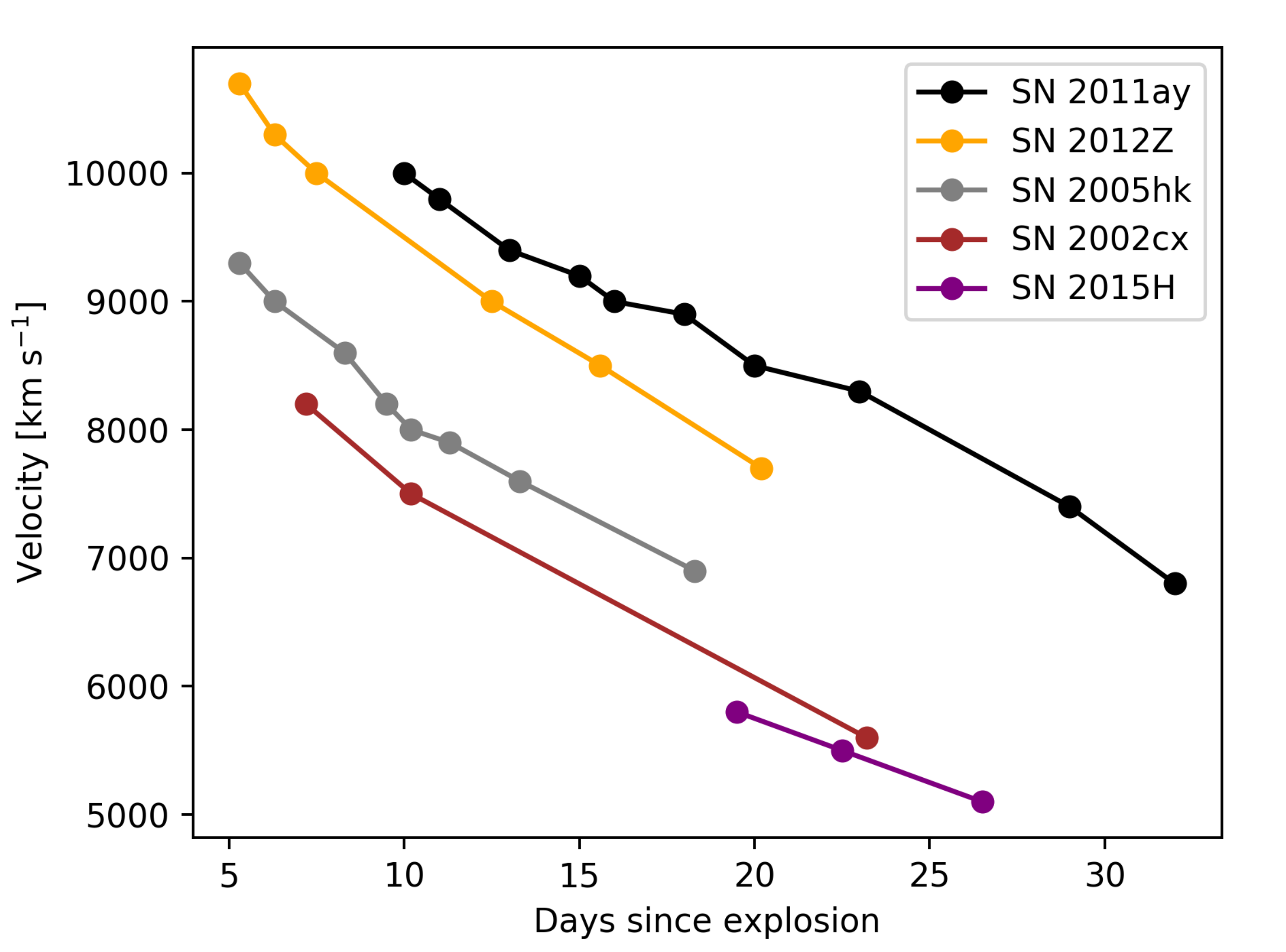}
\caption{\label{fig:velocity} The photospheric velocities as function of time since explosion from our TARDIS fits.}
\end{figure}

The three available spectra for SN 2015H were obtained after maximum light and cover a relatively short time period. Thus, we lose crucial information about the outer regions and can only rely on the line profiles of IGEs and IMEs. Because of higher uncertainties of the chemical profile, we have left out SN 2015H from the construction of the abundance template (see Sec. \ref{results}).

All our model spectra show good matches with the observations of SN 2015H (see Fig. \ref{fig:sn15H_spectra}); some absorption features are reproduced even around $\sim$7\,000 \r{A}. Moreover, the three narrow spectral lines formed by Na\,\begin{small}I\end{small}, Fe\,\begin{small}II\end{small} and Si\,\begin{small}II\end{small} also appear in our synthetic spectra at 5\,890, 5\,914, and 5\,979 \r{A} respectively (see Fig. \ref{fig:sn15H_spectrum_ions}).

\cite{Magee16} fit one spectrum of SN 2015H obtained at +6 days with respect to r-maximum with \begin{small}TARDIS\end{small}. They adopted $T_\rmn{0}$ = 57046.0 MJD as the explosion date, which is only 0.2 days earlier than our result.
The authors used a constant abundance profile, which resulted in a nearly perfect fit using a velocity range of 2\,000 km s$^{-1}$. {\cite{Barna17} showed that abundances in the outer ejecta can have a significant impact on the spectrum. In a broader range of 4\,000 km s$^{-1}$, the stratified abundance profile seems a better solution.} Compared to the photometrically close N3def deflagration model ($M_\rmn{V}$ = -17.52 mag), the similarity of the abundance profile is remarkable under the transition velocity of 7\,000 km s$^{-1}$ as it can be seen in Fig. \ref{fig:ultimate_abundance}. 

%At the same time, the total ejecta mass of our adopted density function, M$_{ej}$ = 0.12 $M_{\odot}$, is significantly lower than the value of M$_{ej}$ = 0.20 $M_{\odot}$ of the N3def model.

Carbon can be allowed in almost the whole volume with a decreasing fraction towards the center, but, again, the uncertainties are higher in the case of SN 2015H.
As for SN 2002cx, we use vanadium in the innermost regions of the modeled volume. The suspected features formed by V\,\begin{small}II\end{small} ($\lambda$3\,951 and $\lambda$4\,883) appear in the spectra obtained at all three epochs. 
%Note that such an uncertain modeling attempt does not mean a definite identification; however, the good match in the late spectra of both SNe 2002cx and 2015H raises its possibility.

\subsection{Evolution of photospheric velocity}
\label{velocity}

{In all the studied SNe, the $v_\rmn{phot}$ values decrease with time monotonically}. The decline rates are higher at the earliest epochs: $\sim$ 200 km s$^{-1}$ day$^{-1}$ at 10 days before maximum light, dropping to $\sim$ 100 km s$^{-1}$ day$^{-1}$ around maximum light, as best seen in Fig. \ref{fig:velocity} for SNe 2005hk and 2012Z. While different supernovae display different velocities, there are no significant differences in the decline rates of the various objects. 

{The $v_\rmn{phot}$ functions (Fig. \ref{fig:velocity}) show a loose correlation  with the peak luminosities in our sample (see Tables \ref{tab:sample} and \ref{tab:log_tardis}) at each epoch; fainter objects tend to have lower expansion velocities.} This might support the idea that a correlation between peak luminosities and expansion velocities exists for the majority of Type Iax SNe. However, there are at least two examples reported in the literature, SN 2014ck \citep{Tomasella16} and SN 2009ku \citep{Narayan11} that show an extremely low photospheric velocity despite having a relatively high peak luminosity. These examples imply that the physical origin of SNe Iax may be more complex so that the whole subclass cannot be described with a few parameters.

The main parameter of the above introduced template abundance profile is the transition velocity. This boundary between the inner regions dominated by IGEs and the outer areas dominated by oxygen appears at a different value in each object. {The transition velocity also shows a loose correlation with the expansion velocities in our sample (see Sec. \ref{sn11ay_results} - \ref{sn2015H_results}). These correlations point in the direction that Type Iax SNe can be described with only a few parameters.}

\section{Conclusions}
\label{conclusions}

We have performed a comprehensive study of the chemical composition and the main physical properties of the diverse Type Iax SN subclass. We analyze the spectral time series of five Type Iax SNe with different peak luminosities and expansion velocities. The data are fit by self-consistent atmosphere models calculated with the spectrum synthesis code \begin{small}TARDIS\end{small}. We follow the modeling strategy introduced in \citet{Barna17} applying a few modifications. We fit the density of the SN ejecta within the frame of density profiles from hydrodynamic deflagration models introducing new free parameters. However, our fits in general prefer a steeper cut-off in  the density profiles at higher velocities than in the deflagration models.

The fitted synthetic spectra show a very good agreement with the observed spectral features. The continuum levels are mostly reproduced well; however, the relatively blue SN 2005hk shows lower continuum flux beyond $\sim$5\,000 \r{A} than our models.

The \begin{small}TARDIS\end{small} model atmospheres are built varying the mass fractions of C, O, Na, Mg, Si, S, Ca, Ti, Cr, Fe and $^{56}$Ni for all five SNe. The starting points of our fitting process are the predicted abundance profiles of the hydrodynamic calculations of pure deflagration models \citep{Fink14}; we deviate from these values only if it is indicated by the fitting of the spectral features. The best-fit models of the two least luminous SNe, 2002ck and 2015H, also include vanadium with mass fractions of 0.01-0.02.

{The most abundant element in our best-fit models is oxygen, just like in the previous abundance tomography made for the spectral series of SN 2011ay \citep{Barna17}. Here we fit the density profiles with a cut-off at higher velocities (Table \ref{tab:log_tardis}) instead of purely exponential profiles. We still have to use oxygen as a  ``filler'' element \citep{Barna17}, despite the reduced mass in the outermost regions of the ejecta.}

Broadly, a similar structure can be recognized in the abundance profiles of the five studied SNe. The trends of the abundance functions are monotonic for each element, no abundance clump is found in the observed sample. Both the mass fractions of IMEs and IGEs decrease towards higher velocities, meanwhile only oxygen and carbon show upward trends. We can place limits only on carbon, which does not appear in the inner ejecta volume (except in case of SN 2015H) studied by the spectral series, its presence is allowed at higher velocities. 
The IGEs are dominated by the products of the $^{56}$Ni decay chain, whose initial mass fraction is typically 0.40-0.50 in the inner regions, while the initial amount of Fe is always below 10 percent. Silicon is one of the best-constrained elements in our analysis because of the sharp Si\,\begin{small}II\end{small} and Si\,\begin{small}III\end{small} lines. The mass fraction reaches 0.10-0.15 at lower velocities and continuously decreases outward. Magnesium, sulfur and calcium show similar trends; their peak values in the innermost regions are $\sim$0.10, $\sim$0.02 and $\sim$0.01, respectively. 

{We have compared our findings to the abundance profiles of deflagration models of similar peak luminosity. We found that our stratified abundance structures deviate from the explosion models significantly at higher velocities and the uniform abundances of the deflagration models do not describe well the outermost part of the Type Iax SN ejecta.}

The mass fractions of the most abundant IMEs and IGEs in our \begin{small}TARDIS\end{small} models are in relatively good agreement with the hydrodynamic calculations in the inner regions of the studied ejecta. However, we can tolerate carbon only in the outermost layers in our spectral fitting, while the deflagration calculations show a constant carbon abundance throughout the ejecta due to turbulent mixing.

{Given that the derived abundance profiles from our model sample show very similar structures, we constructed a template abundance profile. For this purpose, we average the abundances of the individual models in each radial shell with the same relative velocity from the transition velocities (where X($^{56}$Ni) = X(O)). We generated a new set of synthetic spectra shifting the abundance template (Fig. \ref{fig:template_abundance}) in the velocity space according to the different transition velocities (see Sec. \ref{results_template}) of the five individual SN models, . Their fitting with the observed data are nearly on the same level as our original best-fit models. The result that the abundance profile of these objects can be effectively described by one parameter (transition velocity), leads to the conclusion that the origins of the members of the diverse Type Iax subclass may not differ sharply from each other. Moreover, the transition velocities seem to correlate with the peak luminosities and the expansion velocities in our sample. The loose correlation between the peak brightnesses and the expansion velocities has been reported in the literature; although, some outliers have been also observed. Note that, at this point, we cannot classify the whole group of Type Iax SNe as a one-parameter family as the normal SNe Ia. However, we are able to describe the five members of our Type Iax sample with only three parameters: the transition velocity of the abundance profile, the central density, and the location of the cut-off of the density profile. This result points in the direction that these peculiar transient objects can be described with only a few parameters.}

\section*{Acknowledgements}

This work is part of the project Transient Astrophysical Objects GINOP-2-3-2-15-2016-00033 of the National Research, Development and Innovation Office (NK-
FIH), Hungary, funded by the European Union. BB received support by the Campus Mundi Short Study Programme of Tempus Public Foundation through grant no. CM-SMR/252212/2017 and by the New National Excellence Program under UNKP-17-3 grant IV-SZTE-2. TS has received funding from the Hungarian NKFIH/OTKA PD-112325 Grant. WEK and BL acknowledges the Excellence Cluster Universe, Technische Universit\"at M\"unchen, Boltzmannstrasse 2, D-85748 Garching, Germany and WEK acknowledges the support of an ESO Fellowship. MK acknowledges support from the Klaus Tschira Foundation. SAS acknowledges support from STFC via grant ST/P000312/1. 
This study has made use of The Weizmann interactive supernova data repository - http://wiserep.weizmann.ac.il - and the Open Supernova Catalog.

%%%%%%%%%%%%%%%%%%%%%%%%%%%%%%%%%%%%%%%%%%%%%%%%%%



%\bibliographystyle{mnras}


%%%%%%%%%%%%%%%%%%%%%%%%%%%%%%%%%%%%%%%%%%%%%%%%%%

%%%%%%%%%%%%%%%%% APPENDICES %%%%%%%%%%%%%%%%%%%%%

\appendix

\section{Abundance parameters of the best-fit models}

\begin{table*}
	\centering
	\caption{Mass fractions of chemical elements in velocity layers (between v$_\rmn{low}$ and v$_\rmn{up}$ in km s$^{-1}$) of the best-fit model for SN 2011ay.}
    \label{tab:data_sn11ay}
    \begin{tabular}{ccccccccccccc}
		\hline
		X(C) & X(O) & X(Na) & X(Mg) & X(Si) & X(S) & X(Ca) & X(Ti) & X(Cr) & X(Fe) & X($^{56}$Ni) & v$_\rmn{low}$ & v$_\rmn{up}$\\
		\hline
0.000 & 0.150 & 0.000 & 0.100 & 0.150 & 0.030 & 0.005 & 0.020 & 0.020 & 0.100 & 0.440 & 7000 & 7500\\
0.000 & 0.155 & 0.000 & 0.100 & 0.150 & 0.025 & 0.005 & 0.020 & 0.020 & 0.100 & 0.440 & 7500 & 8000\\
0.000 & 0.190 & 0.000 & 0.100 & 0.140 & 0.015 & 0.005 & 0.010 & 0.015 & 0.100 & 0.440 & 8000 & 8500\\
0.000 & 0.245 & 0.000 & 0.100 & 0.130 & 0.010 & 0.003 & 0.010 & 0.015 & 0.080 & 0.420 & 8500 & 9000\\
0.000 & 0.270 & 0.000 & 0.090 & 0.110 & 0.010 & 0.003 & 0.005 & 0.015 & 0.080 & 0.420 & 9000 & 9500\\
0.000 & 0.315 & 0.000 & 0.070 & 0.090 & 0.010 & 0.003 & 0.000 & 0.010 & 0.080 & 0.420 & 9500 & 10000\\
0.000 & 0.345 & 0.030 & 0.050 & 0.075 & 0.010 & 0.003 & 0.000 & 0.000 & 0.070 & 0.420 & 10000 & 10500\\
0.000 & 0.370 & 0.050 & 0.040 & 0.050 & 0.010 & 0.002 & 0.000 & 0.000 & 0.060 & 0.420 & 10500 & 11000\\
0.000 & 0.475 & 0.050 & 0.020 & 0.030 & 0.005 & 0.001 & 0.000 & 0.000 & 0.050 & 0.370 & 11000 & 11500\\
0.040 & 0.580 & 0.050 & 0.000 & 0.020 & 0.000 & 0.000 & 0.000 & 0.000 & 0.030 & 0.280 & 11500 & 12000\\
0.100 & 0.670 & 0.050 & 0.000 & 0.000 & 0.000 & 0.000 & 0.000 & 0.000 & 0.000 & 0.180 & 12000 & 13000\\
0.100 & 0.750 & 0.050 & 0.000 & 0.000 & 0.000 & 0.000 & 0.000 & 0.000 & 0.000 & 0.100 & 13000 & 14000\\
0.150 & 0.800 & 0.050 & 0.000 & 0.000 & 0.000 & 0.000 & 0.000 & 0.000 & 0.000 & 0.000 & 14000 & 15000\\
        \hline
	\end{tabular}
\end{table*}

\begin{table*}
	\centering
	\caption{Mass fractions of chemical elements in velocity layers (between v$_\rmn{low}$ and v$_\rmn{up}$ in km s$^{-1}$) of the best-fit model for SN 2012Z.}
    \label{tab:data_sn12Z}
    \begin{tabular}{ccccccccccccc}
		\hline
		X(C) & X(O) & X(Na) & X(Mg) & X(Si) & X(S) & X(Ca) & X(Ti) & X(Cr) & X(Fe) & X($^{56}$Ni) & v$_\rmn{low}$ & v$_\rmn{up}$\\
		\hline
0.000 & 0.130 & 0.000 & 0.090 & 0.120 & 0.025 & 0.003 & 0.010 & 0.020 & 0.100 & 0.500 & 7500 & 8000\\
0.000 & 0.140 & 0.000 & 0.090 & 0.120 & 0.015 & 0.003 & 0.010 & 0.020 & 0.100 & 0.500 & 8000 & 8500\\
0.000 & 0.140 & 0.000 & 0.090 & 0.120 & 0.015 & 0.003 & 0.010 & 0.020 & 0.100 & 0.500 & 8500 & 9000\\
0.000 & 0.150 & 0.000 & 0.090 & 0.120 & 0.005 & 0.003 & 0.010 & 0.020 & 0.100 & 0.500 & 9000 & 9500\\
0.000 & 0.185 & 0.000 & 0.090 & 0.110 & 0.005 & 0.003 & 0.010 & 0.015 & 0.090 & 0.490 & 9500 & 10000\\
0.000 & 0.225 & 0.000 & 0.080 & 0.100 & 0.005 & 0.003 & 0.005 & 0.005 & 0.085 & 0.480 & 10000 & 10500\\
0.000 & 0.280 & 0.000 & 0.080 & 0.090 & 0.005 & 0.003 & 0.000 & 0.000 & 0.080 & 0.460 & 10500 & 11000\\
0.020 & 0.380 & 0.030 & 0.080 & 0.060 & 0.000 & 0.002 & 0.000 & 0.000 & 0.070 & 0.360 & 11000 & 12000\\
0.100 & 0.430 & 0.050 & 0.050 & 0.020 & 0.000 & 0.001 & 0.000 & 0.000 & 0.050 & 0.300 & 12000 & 13000\\
0.170 & 0.550 & 0.050 & 0.030 & 0.000 & 0.000 & 0.000 & 0.000 & 0.000 & 0.000 & 0.200 & 13000 & 14000\\
0.250 & 0.700 & 0.050 & 0.000 & 0.000 & 0.000 & 0.000 & 0.000 & 0.000 & 0.000 & 0.000 & 14000 & 15000\\
0.250 & 0.700 & 0.050 & 0.000 & 0.000 & 0.000 & 0.000 & 0.000 & 0.000 & 0.000 & 0.000 & 15000 & 16000\\ 
        \hline
	\end{tabular}
\end{table*}

\begin{table*}
	\centering
	\caption{Mass fractions of chemical elements in velocity layers (between v$_\rmn{low}$ and v$_\rmn{up}$ in km s$^{-1}$) of the best-fit model for SN 2005hk.}
    \label{tab:data_sn05hk}
    \begin{tabular}{ccccccccccccc}
		\hline
		X(C) & X(O) & X(Na) & X(Mg) & X(Si) & X(S) & X(Ca) & X(Ti) & X(Cr) & X(Fe) & X($^{56}$Ni) & v$_\rmn{low}$ & v$_\rmn{up}$\\
		\hline
0.000 & 0.115 & 0.000 & 0.100 & 0.155 & 0.010 & 0.005 & 0.015 & 0.020 & 0.080 & 0.500 & 7000 & 7500\\
0.000 & 0.140 & 0.000 & 0.100 & 0.150 & 0.010 & 0.005 & 0.015 & 0.015 & 0.080 & 0.500 & 7500 & 8000\\
0.000 & 0.140 & 0.000 & 0.100 & 0.150 & 0.010 & 0.005 & 0.010 & 0.005 & 0.080 & 0.500 & 8000 & 8500\\
0.000 & 0.180 & 0.000 & 0.095 & 0.120 & 0.005 & 0.005 & 0.010 & 0.005 & 0.080 & 0.500 & 8500 & 9000\\
0.000 & 0.200 & 0.030 & 0.095 & 0.090 & 0.000 & 0.005 & 0.000 & 0.000 & 0.080 & 0.500 & 9000 & 9500\\
0.040 & 0.245 & 0.050 & 0.075 & 0.070 & 0.000 & 0.003 & 0.000 & 0.000 & 0.070 & 0.450 & 9500 & 10000\\
0.090 & 0.370 & 0.050 & 0.060 & 0.050 & 0.000 & 0.001 & 0.000 & 0.000 & 0.050 & 0.320 & 10000 & 10500\\
0.100 & 0.500 & 0.050 & 0.050 & 0.020 & 0.000 & 0.000 & 0.000 & 0.000 & 0.000 & 0.280 & 10500 & 11000\\
0.200 & 0.570 & 0.050 & 0.000 & 0.000 & 0.000 & 0.000 & 0.000 & 0.000 & 0.000 & 0.180 & 11000 & 12000\\
0.250 & 0.700 & 0.050 & 0.000 & 0.000 & 0.000 & 0.000 & 0.000 & 0.000 & 0.000 & 0.000 & 12000 & 13000\\
0.250 & 0.700 & 0.050 & 0.000 & 0.000 & 0.000 & 0.000 & 0.000 & 0.000 & 0.000 & 0.000 & 13000 & 14000\\
        \hline
	\end{tabular}
\end{table*}

\begin{table*}
	\centering
	\caption{Mass fractions of chemical elements in velocity layers (between v$_\rmn{low}$ and v$_\rmn{up}$ in km s$^{-1}$) of the best-fit model for SN 2002cx.}
    \label{tab:data_sn02cx}
    \begin{tabular}{cccccccccccccc}
		\hline
		X(C) & X(O) & X(Na) & X(Mg) & X(Si) & X(S) & X(Ca) & X(Ti) & X(V) & X(Cr) & X(Fe) & X($^{56}$Ni) & v$_\rmn{low}$ & v$_\rmn{up}$\\
		\hline
0.000 & 0.170 & 0.000 & 0.080 & 0.150 & 0.020 & 0.005 & 0.005 & 0.020 & 0.020 & 0.070 & 0.460 & 5000 & 5500\\
0.000 & 0.200 & 0.000 & 0.080 & 0.120 & 0.020 & 0.005 & 0.005 & 0.020 & 0.020 & 0.070 & 0.460 & 5500 & 6000\\
0.000 & 0.230 & 0.000 & 0.080 & 0.100 & 0.020 & 0.005 & 0.005 & 0.015 & 0.015 & 0.070 & 0.460 & 6000 & 6500\\
0.000 & 0.270 & 0.000 & 0.080 & 0.080 & 0.020 & 0.005 & 0.005 & 0.015 & 0.015 & 0.070 & 0.440 & 6500 & 7000\\
0.000 & 0.310 & 0.000 & 0.080 & 0.070 & 0.015 & 0.005 & 0.000 & 0.010 & 0.010 & 0.070 & 0.430 & 7000 & 7500\\
0.000 & 0.330 & 0.000 & 0.080 & 0.070 & 0.015 & 0.005 & 0.000 & 0.000 & 0.000 & 0.070 & 0.430 & 7500 & 8000\\
0.000 & 0.350 & 0.005 & 0.080 & 0.030 & 0.010 & 0.005 & 0.000 & 0.000 & 0.000 & 0.070 & 0.440 & 8000 & 8500\\
0.000 & 0.470 & 0.030 & 0.070 & 0.000 & 0.000 & 0.005 & 0.000 & 0.000 & 0.000 & 0.055 & 0.370 & 8500 & 9000\\
0.060 & 0.520 & 0.050 & 0.030 & 0.000 & 0.000 & 0.004 & 0.000 & 0.000 & 0.000 & 0.040 & 0.300 & 9000 & 9500\\
0.150 & 0.620 & 0.050 & 0.000 & 0.000 & 0.000 & 0.002 & 0.000 & 0.000 & 0.000 & 0.020 & 0.160 & 9500 & 10500\\
0.200 & 0.700 & 0.050 & 0.000 & 0.000 & 0.000 & 0.000 & 0.000 & 0.000 & 0.000 & 0.000 & 0.000 & 10500 & 11500\\
0.250 & 0.700 & 0.050 & 0.000 & 0.000 & 0.000 & 0.000 & 0.000 & 0.000 & 0.000 & 0.000 & 0.000 & 11500 & 12500\\
0.250 & 0.700 & 0.050 & 0.000 & 0.000 & 0.000 & 0.000 & 0.000 & 0.000 & 0.000 & 0.000 & 0.000 & 12500 & 13500\\
        \hline
	\end{tabular}
\end{table*}

\begin{table*}
	\centering
	\caption{Mass fractions of chemical elements in velocity layers (between v$_\rmn{low}$ and v$_\rmn{up}$ in km s$^{-1}$) of the best-fit model for SN 2015H.}
    \label{tab:data_sn15H}
    \begin{tabular}{cccccccccccccc}
		\hline
		X(C) & X(O) & X(Na) & X(Mg) & X(Si) & X(S) & X(Ca) & X(Ti) & X(V) & X(Cr) & X(Fe) & X($^{56}$Ni) & v$_\rmn{low}$ & v$_\rmn{up}$\\
		\hline
0.050 & 0.160 & 0.000 & 0.080 & 0.140 & 0.025 & 0.005 & 0.000 & 0.020 & 0.020 & 0.080 & 0.400 & 5000 & 5500\\
0.080 & 0.190 & 0.000 & 0.080 & 0.140 & 0.025 & 0.005 & 0.000 & 0.010 & 0.020 & 0.080 & 0.370 & 5500 & 6000\\
0.120 & 0.240 & 0.000 & 0.080 & 0.085 & 0.020 & 0.005 & 0.000 & 0.000 & 0.020 & 0.080 & 0.350 & 6000 & 6500\\
0.170 & 0.280 & 0.000 & 0.070 & 0.025 & 0.010 & 0.004 & 0.000 & 0.000 & 0.010 & 0.080 & 0.350 & 6500 & 7000\\
0.180 & 0.380 & 0.050 & 0.055 & 0.000 & 0.000 & 0.004 & 0.000 & 0.000 & 0.000 & 0.060 & 0.280 & 7000 & 7500\\
0.200 & 0.530 & 0.050 & 0.050 & 0.000 & 0.000 & 0.003 & 0.000 & 0.000 & 0.000 & 0.050 & 0.120 & 7500 & 8000\\
0.250 & 0.650 & 0.050 & 0.050 & 0.000 & 0.000 & 0.001 & 0.000 & 0.000 & 0.000 & 0.000 & 0.000 & 8000 & 8500\\
0.250 & 0.650 & 0.050 & 0.050 & 0.000 & 0.000 & 0.000 & 0.000 & 0.000 & 0.000 & 0.000 & 0.000 & 8500 & 9000\\
0.300 & 0.650 & 0.050 & 0.000 & 0.000 & 0.000 & 0.000 & 0.000 & 0.000 & 0.000 & 0.000 & 0.000 & 9000 & 9500\\
0.300 & 0.650 & 0.050 & 0.000 & 0.000 & 0.000 & 0.000 & 0.000 & 0.000 & 0.000 & 0.000 & 0.000 & 9500 & 10000\\
        \hline
	\end{tabular}
\end{table*}

%%%%%%%%%%%%%%%%%%%%%%%%%%%%%%%%%%%%%%%%%%%%%%%%%%

% Don't change these lines
\bsp	% typesetting comment
\label{lastpage}
\end{document}